\documentclass[article,3p,times,onecolumn]{elsarticle}
\usepackage{lineno,hyperref}
\modulolinenumbers[5]
\usepackage{comment}
\usepackage{latexsym}
\usepackage{epsfig}
\usepackage{amsmath}
\usepackage{amsfonts}
\usepackage{amssymb}
\usepackage{amsthm}
\usepackage{graphicx}
\usepackage{wrapfig}
\usepackage{pgf}
\usepackage{caption}
\usepackage{subcaption}
\usepackage{natbib}
\usepackage{tikz}
\usepackage[font=normalsize]{caption}

\graphicspath{{Figures/}}

\journal{J.~non-Newt.~Fluid Mech.}

\begin{document}
	
	\begin{frontmatter}
		
		\medskip  
		
		\title{Turbulent drag reduction of viscoelastic wormlike micellar gels}
		
		\medskip
		
		\author[1]{Rodrigo~S.~Mitishita\corref{cor1}} 
		\ead{rodrigo.seiji06@gmail.com} 
		\cortext[cor1]{Corresponding author}
		
		\author[1]{Gwynn~J.~Elfring} 
		
		\author[1,2]{Ian.~A.~Frigaard}
		
		\address[1]{Department of Mechanical Engineering, University of British Columbia, 2324 Main Mall, Vancouver, BC, V6T 1Z4, Canada}
		
		\address[2]{Department of Mathematics, University of British Columbia, 1984 Mathematics Rd, Vancouver, BC, V6T 1Z2, Canada}

		\begin{abstract}
			Long-chained, viscoelastic surfactant solutions (VES) have been widely employed in the oil and gas industry, particularly in hydraulic fracturing and gravel-packing operations, where turbulence is commonly reached due to high pumping rates. With this motivation, we experimentally investigate the turbulent duct flow of an under-studied class of wormlike micellar solutions that forms a gel at room temperature. The fluid is characterized via rotational rheometry, and the turbulent velocity and Reynolds stress profiles are measured via Laser Doppler Anemometry (LDA). Three surfactant concentrations are investigated at increasing Reynolds numbers. The turbulent flow fields of water, and semi-dilute solutions of partially hydrolyzed polyacrylamide (HPAM) and xanthan gum (XG) are used as comparisons. Our study reveals that the gel-like structure of the wormlike micellar gel is mostly broken down during turbulent flow, especially in the near-wall region where the results indicate the presence of a water layer. Turbulent flow at low concentrations of surfactant show a Newtonian-like flow field throughout most of the duct, where the energy spectra shows a $-5/3$ power law scale with wavenumber, whereas higher concentrations lead to drag reduction and lower power spectral densities at large wavenumbers. A comparison of the flow of polymeric fluids and the wormlike micellar solutions at maximum drag reduction (MDR) shows comparable drag-reduction effects, with a large decrease in Reynolds shear stresses, and increased turbulence anisotropy in the buffer layer due to the large streamwise fluctuations and near-zero wall-normal fluctuations. At the centreline of the duct, the power spectral densities of polymers and VES are significantly reduced at all wavenumbers probed. Additionally, the MDR regime was bounded by Virk's asymptote for both polymers and the micellar gel, which implies a similar mechanism of drag reduction at MDR.
			
		\end{abstract}
		
		
	\end{frontmatter}
	
	\section{Introduction} \label{s1}
	\label{sec:intro}
	
	Amphiphilic surfactant molecules in aqueous solutions can assemble into a variety of structures such as spheres, rods, long entangled ``worms" (hence the name wormlike micelles) and branched micelles as the surfactant concentration increases \citep{cates1990b, zakin1998}. Aside from the concentration changes, the micellar morphology can be significantly affected by changes in temperature, the presence of salt in the solution \citep{cates1990b, raghavan2001} or charges carried by surfactant molecules \citep{zakin1998}. The rheological behaviour of micellar fluids shares some similarities with polymer solutions \citep{cates1990c}, but while polymers can be permanently degraded by shear or extensional forces due to chain scission, micelles are able to dynamically break and re-build their structure. Surfactants are commonly employed in many commercial applications such as hydraulic fracturing \citep{barbati2016}, personal care \citep{yang2002} and cleaning, but the main focus of this paper is in turbulent drag reduction (TDR) applications for surfactant solutions, specifically for wormlike micellar fluids in the semi-dilute regime. TDR has been mostly investigated with dilute polymer solutions, but surfactant solutions have also been favoured in recirculating applications due to recoverable micellar structure \citep{zakin1998}.
	
	Since the first TDR experiments by \citet{toms1977}, it has been well known that the addition of polymers to a turbulent flow of water can decrease frictional drag by over 70\% \citep{virk1970, escudier2009a}. This discovery led to significant energy savings when pumping fluids across long distances such as crude oil transport \citep{white2008} and fire fighting operations \citep{pereira2012}. Despite being a subject investigated for more than four decades and subject of many comprehensive reviews such as \citep{virk1975,white2008,graham2014,xi2019}, the turbulent drag reduction (TDR) phenomenon remains a subject of interest for researchers, and in particular much less is known about TDR in surfactant fluids. 
	
	The main industrial motivation for our work is from the oil and gas industry, specifically a well completion procedure for unconsolidated reservoirs called gravel-packing. Polymer and surfactant fluids are employed to suspend solid particles that are pumped downhole through rectangular aspect ratio ducts called shunt tubes, typically across hundreds of metres. Even with the relatively high viscosities required to suspend the particles, the fluid eventually becomes turbulent in the shunt tubes due to the high flow velocities. Turbulent drag reduction is then possible due to the viscoelastic properties of polymer and wormlike micellar solutions, greatly reducing the pumping power required. The particles are then placed between the reservoir and a mesh screen to act as a filter to sand particles and to add structural rigidity the well \citep{shirazi2020}. Examples of gravel-packing fluids are xanthan gum and zwitterionic surfactants in semi-dilute concentrations \citep{jain2011,goyal2017}. In this paper, we put forward an experimental investigation of the turbulent flow of one such surfactant solution, which was recently characterized via shear rheology for the most part \citep{kumar2007,beaumont2013,wang2017,gupta2021}, but whose high Reynolds number turbulent characteristics are relatively unknown.
	
	\subsection{Polymer drag reduction} \label{s1.1}
	
	Theoretical studies of TDR with polymer solutions have outlined two major explanations for the drag reduction mechanism: the viscous theory by \citet{lumley1969} and the elastic theory by \citet{tabor1986}. The viscous theory suggests that the presence of polymer causes an increase in viscosity near the wall due to extension of the molecules, which suppresses turbulent fluctuations and leads to drag reduction. The elastic theory by \citet{tabor1986} proposes that elastic stresses that arise from the stretching of the polymers become of the same order as the Reynolds stresses, but at a length scale larger than the Kolmogorov microscale. The elastic stresses then suppress the smaller eddies, cutting off the energy cascade and reducing the frictional drag.
	
	The behaviour of polymeric fluids in turbulent flows has been extensively investigated over the years via experiments (\citep{lumley1969, virk1970, warholic2001, white2004, pereira2013, shaban2018, choueiri2018}, among many others). When considering flexible polymeric additives, the consensus is that viscoelasticity plays an important role in weakening turbulence in drag-reducing flows. The effect of viscoelasticity in a duct or pipe flow is usually quantified by the Weissenberg number $Wi$, which is the product of the extensional relaxation time of the viscoelastic fluid (usually measured with an extensional rheometer such as the Capillary Breakup Extensional Rheometer, or CaBER) and the characteristic shear rate of the flow, given as the ratio between the average velocity and hydraulic diameter. With higher $Wi$, the time-averaged flow velocity increases, while and the friction factor and the Reynolds shear stresses decrease, indicating a direct relationship between viscoelasticity and TDR \citep{white2008}. Recently, \citet{owolabi2017} was able to experimentally correlate the Weissenberg number of PAM solutions to drag reduction. Experimental and numerical studies of TDR also uncovered the decrease in ejection and sweep events \citep{shaban2018}, and the role played by elastic stresses in dampening vortical motions as energy is transferred from the eddies to polymer or surfactant molecules \citep{warholic2001, pereira2017b}. Advancements in the field of polymer drag reduction have been consequence of direct numerical simulations (DNS), usually with the finitely extensible nonlinear elastic dumbbell model with the Peterlin approximation (FENE-P) model. DNS are able to resolve the flow up to the small (Kolmogorov) scales, and allow the analysis of data that are difficult to obtain experimentally, such as three components of velocity and vorticity and visualization of vortex structures \citep{pereira2017b, xi2010}. 

	Drag reduction increases with $Wi$ or polymer/surfactant concentration until the point of maximum drag reduction (MDR), where further increase in concentration does not contribute to drag reduction \citep{virk1970}.  Both numerical simulations \citep{xi2012b, dubief2013, graham2014} and experiments \citep{samanta2013, choueiri2018} have contributed in explaining the mechanism of the MDR asymptote. Intermittent intervals of active turbulence stretch the polymer molecules, leading to a hibernating state of low turbulence intensity where the polymer molecules return to their initial coiled state \citep{xi2012a, pereira2017c}. More recently, the interaction between inertial and elastic instabilities in a novel turbulence state named elasto-inertial turbulence \citep{dubief2013, zhu2021} has been proposed as a pathway to MDR. 
	
	A large amount of the numerical and experimental studies on drag reduction have focused on flexible polymer solutions given the correlation of TDR and $Wi$, and with the polymer molecules simplified as elastic dumbells in the FENE-P model, for instance. Thus, rigid polymer solutions such as xanthan gum (XG) have received less attention until recently \citep{xi2019}. Polymer drag reduction  with flexible molecules is usually of type A, characterized by a sudden ``onset" of drag reduction (or sudden decrease in the friction factor) at a critical Reynolds number ($Re$), with an increase in TDR with larger concentration and $Re$ values until the MDR state is reached \citep{andrade2014}. Conversely, the type B drag reduction manifests itself as mostly concentration dependent, and increasing $Re$ does not result in large changes in the drag reduction \citep{virk1990, jaafar2009, gasljevic2001, pereira2013}. In addition to their flow phenomenology, flexible and rigid polymeric additives have been compared in terms of their resistance to degradation. \citet{pereira2013} observed that xanthan gum (rigid polymer) appears to form aggregates as concentration increases, and its degradation in turbulent flow could be a destruction of these aggregates in shear, instead of suffering permanent chain scission like flexible polymers such as poly(ethylene oxide) (PEO). The Weissenberg number has been shown to correlate to drag reduction in flexible polymers, but the picture is less clear when rigid polymers are added. The relaxation time of rigid polymers was observed to be much lower than flexible polymers \citep{pereira2013}, which indicates that the $Wi$ may not be an adequate quantity to gauge drag reduction in all polymer flows \citep{pereira2013, mohammadtabar2020, warwaruk2021}. Thus, it appears that it may not be possible to correlate TDR to a single rheological property of polymeric fluids.
	
	\subsection{Surfactant drag reduction} \label{s1.2}
	
	Differently from polymeric fluids, drag reduction for dilute surfactant fluids has been linked to the appearance of shear induced structures (SIS) in a turbulent flow, caused by alignment and subsequent aggregatopm of rodlike micelles under shear \citep{protzl1997} up to a few microns in length. The type and concentration of salt in solution, temperature and surfactant concentration also influence whether or not SIS formation happens in dilute surfactant solutions. The presence of SIS can be probed in shear-rate-controlled rheometer experiments, where it is correlated to shear-thickening viscosity measurements \citep{drappier2006, vasudevan2008}. Shear-thickening in steady flow can also be correlated to observations in light scattering \citep{liu1996, protzl1997} and small-angle neutron scattering investigations \citep{takeda2011}, where anisotropic scattering patterns oriented in the flow direction are visualized simultaneously with an increase in viscosity measurements.
	
	SIS have been correlated to drag reduction in a few recent experimental investigations. Evidence of SIS have been observed via particle image velocimetry (PIV) and direct imaging in the experiments in a pipe flow \citep{tuan2013} and channel flow in \citep{tuan2017} with a shear-thickening surfactant solution, where the SIS appeared as turbid, light-scattering threads in the pipe. Their observations have been correlated to friction factor \textit{vs.} Reynolds number measurements, where the SIS threads were visualized alongside decreased friction factor observations (TDR), and disappeared as the friction factor increased due to degradation. Similar gel-like threads were observed in a Couette flow of surfactants by \citet{liu1996}, which imparts viscoelastic characteristics to the surfactant fluids, which enables turbulent drag reduction \citep{tamano2009}.
	
	Recently, \citet{wakimoto2018} investigated TDR with a shear-thickening surfactant fluid via simultaneous pressure/flow rate measurements and fluorescence probing, where the fluorescence intensity is anti-correlated to the presence of SIS. This occurs due to the fact that fluorescent molecules are trapped within the SIS, thus decreasing the measured intensity. Their results suggest that SIS are formed near the pipe wall, but are broken down once the shear rate becomes to high, resulting in degradation and loss of drag reduction. However, not all drag-reducing surfactant solutions are shear thickening and form SIS. In semi-dilute or higher concentrations, shear-thinning viscosities are observed during a steady-shear flow curve, also a characteristic of the rheology of drag-reducing polymer solutions \citep{warholic1999,aguilar2001,haward2012}. The experimental investigation from \citet{warholic1999}, for example, investigated a semi-dilute, shear-thinning surfactant solution at MDR. Their turbulence quantities were close to what is observed in polymers at MDR \citep{escudier2009a, warholic1999influence, owolabi2017}, with near zero Reynolds shear stress profiles and velocity profiles matching Virk's asymptote \citep{virk1970}. The experimental data led the authors to conclude that the entangled wormlike micelles interact with and dampen turbulent structures, similarly to how the SIS interact with turbulent flows.
	
	\subsection{Wormlike micellar gels} \label{s1.3}
	
	The surfactants of interest in the present study are shear-thinning, long chained (C22, or systems with 22 carbon atom tails) surfactant solutions, which encompass an understudied class of wormlike micellar fluids, at least in the context of turbulent drag reduction, with the majority of investigations being from a rheological perspective. \citet{raghavan2001} presented a rheological study of C22-tailed surfactants named EHAC and ETAC. Measurements of the viscoelastic moduli in the linear viscoelastic regime revealed a gel-like rheology with very long relaxation times at room temperature, and Maxwell-like rheology at high temperatures, along with shear-thinning rheology and lack of SIS formation. Similar results were presented for an aqueous solution of erucyl dimethyl amidopropyl betaine (EDAB), also a C22 surfactant \citep{kumar2007}. The gel-like rheology of EDAB also correlated with observations of an apparent yield stress, which is a quite unique observation in micellar solutions, since those were mainly observed to be shear-thinning, viscoelastic fluids with a Maxwell rheology in the linear regime \citep{cates2006rheology}. The yield stress is a consequence of entanglement of very long wormlike micelles, evidenced by cryo-TEM visualizations \citep{kumar2007, raghavan2012}. Furthermore, the rheological properties of EDAB are insensitive to salt additions, since it is a zwitterionic surfactant. Additional rheometer and Couette flow visualizations revealed high viscoelasticity via measurements of large first normal stress differences $N_1$ in steady shear, and evidence of elastic turbulence in Couette flow at low Reynolds numbers. These elastic instabilities are a consequence of large $Wi$ values due to the very long relaxation times \citep{beaumont2013}.
	
	The focus of our experiments is a viscoelastic surfactant (VES) similar to an EDAB solution, which was subject of a recent rheological study by \citet{gupta2021}. In addition to the aforementioned rheological studies, turbulent flow investigations have been rare, with the only published data being from \citet{jain2011} and \citet{goyal2017}. However, they only presented friction factor data as a function of the Reynolds number with limited rheological measurements. So far, a detailed analysis of the turbulent flow field of an EDAB or similar solution has not yet been performed, and the similarities or differences to better known polymer or surfactant solutions are mostly unknown, to the best of our knowledge. To account for this gap in the literature, the objective of this paper is to present a comprehensive turbulent flow investigation of VES/EDAB via high-resolution LDA measurements of the velocity and Reynolds stress field, with further measurements of probability density functions and power spectra of velocity fluctuations with EDAB solutions at three concentrations in the semi-dilute regime. The VES results are compared to turbulent quantities of water, HPAM and XG solutions (flexible and rigid polymers, respectively), to possibly discern the DR mechanisms of polymers and the gel-like surfactant.
	
	\subsection{Outline} \label{s1.4}
	
	The paper is organized as follows. In \hyperref[s2]{Section~\ref*{s2}} we outline the experimental setup, fluid preparation and LDA and rheometry procedures. We discuss the results from the rheological characterization of the fluids in \hyperref[s3]{Section~\ref*{s3}}. The main results from the turbulent velocity measurements are presented in \hyperref[s4]{Section~\ref*{s4}}, where the effect of concentration of VES is assessed, followed by a comparison between the MDR states of the VES fluid and polymer solutions. Finally, the conclusion and summation of the main findings are presented in \hyperref[s6]{Section~\ref*{s6}}.
	
	\section{Experiments} \label{s2}
	
	\subsection{Flow loop} \label{s2.1}
	
\begin{figure}
		\centering
		\includegraphics*[width=140mm]{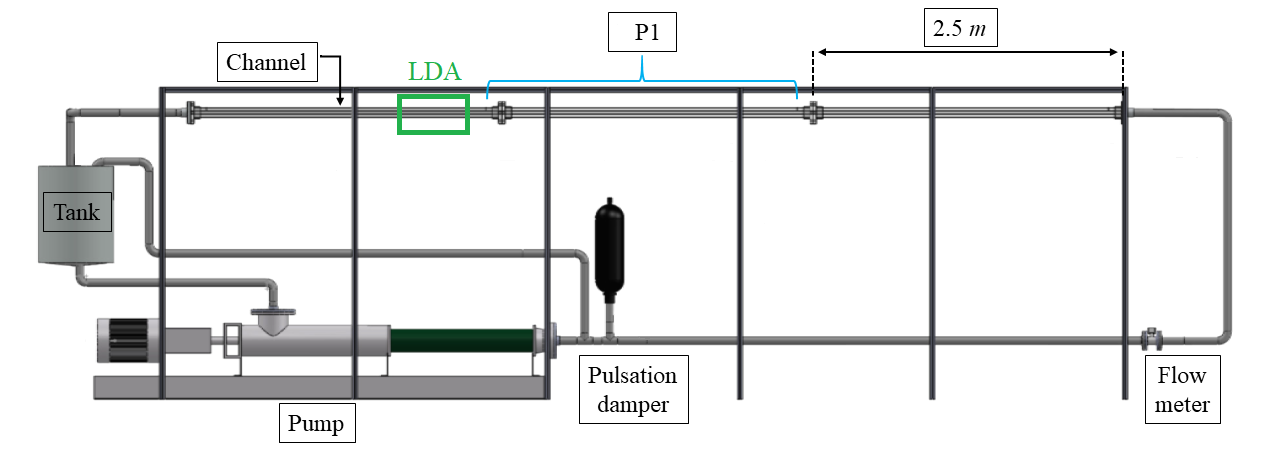}
		\caption{\fontsize{9}{9}\selectfont Flow loop schematic.}
		\label{flowloop1}
	\end{figure}

\begin{figure}
		\centering
		\includegraphics*[width=120mm]{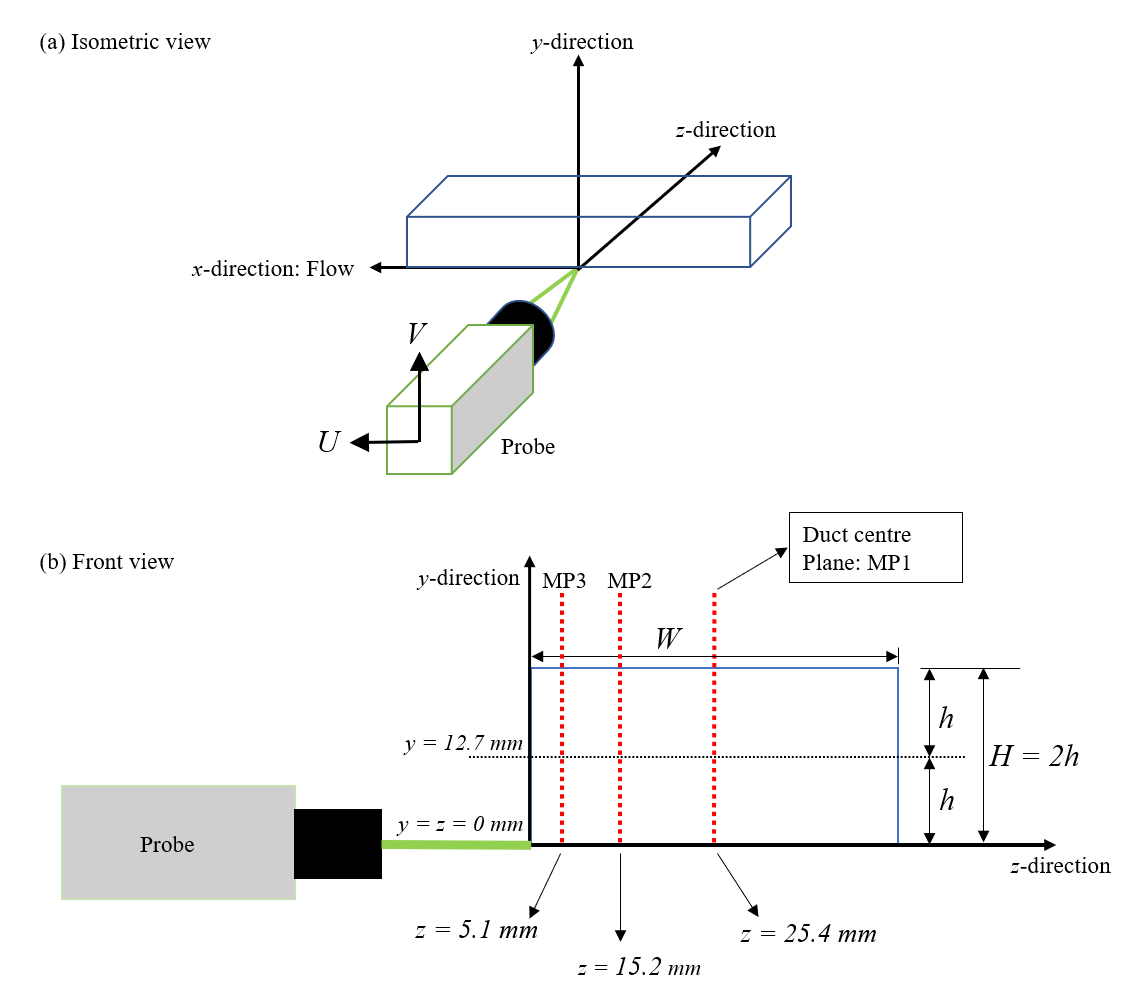}
		\caption{\fontsize{9}{9}\selectfont Flow loop schematic.}
		\label{flowloop2}
	\end{figure}

	We carry out our turbulent flow investigation in a horizontal, pump driven flow loop, which is described in detail in our previous work \citep{mitishita2021}, and thus only a summary of the setup is provided here. The test section is a transparent $7.5~m$ duct with a rectangular cross section, connected to the pump discharge pipeline by directional valves. A schematic of the flow loop is presented in \hyperref[flowloop1]{Figure~\ref*{flowloop1}}. The rectangular test section (``Duct'' in \hyperref[flowloop1]{Figure~\ref*{flowloop1}}) is made of three 2.5 $m$ long, clear acrylic channels connected together by flanges. The inner dimensions of the duct are width of $W = 50.8~mm  = 2w$ and height of $H = 25.4~mm = 2h$, where $w$ is the channel half-width and $h$ is the channel half-height, with hydraulic diameter $D_h = 2WH/(W+H)$ of $33.8~mm$. The test section is presented in more detail in the schematic of \hyperref[flowloop2]{Figure~\ref*{flowloop2}}. The flow is driven by a Netzsch NEMO progressing cavity pump connected to a variable frequency drive, capable of a maximum flow rate of 1200 $l/min$. A Parker pulsation damper was installed after the pump discharge to prevent pressure fluctuations. An Omega FMG 606 magnetic flow meter, with accuracy of 0.5\% of full scale is used to measure the average flow rate. The pressure drop along 2.5 \textit{m} of the rectangular test section is measured by an Omega DPG 409 differential pressure transducer (water and polymer experiments), with 0.08 \% accuracy relative to the full range of 0 to 50 \textit{psi} and data acquisition rate of 3 \textit{hz}. We also employed a PX419 pressure transducer for the VES experiments, with 0.08\% accuracy relative to the full range of 0 to 15 \textit{psi} and data acquisition rate of 3 \textit{hz}. The upstream and downstream ports (marked by P1 in \hyperref[flowloop1]{Figure~\ref*{flowloop1}}) were connected to the test section by hoses filled with water, attached to  pressure taps of 3\textit{mm} diameter on the top wall of the duct aligned with the centreline. The temperature of the fluid in the tank is measured by an Omega TC-NPT pipe thermocouple, of 0.5\% accuracy. Remote control of the pump speed, as well as the signal acquisition from the thermocouple, pressure transducers and flow meter are provided by the National Instruments LabVIEW software and compact data acquisition modules.
	
	We perform velocity measurements with a two-component Laser Doppler Anemometer (LDA) from Dantec Dynamics in backscatter mode at over 5 \textit{m} (or approximately 150$D_h$) downstream of the test section inlet, where we consider the flow to be fully developed. The receiving optics and laser system are contained in a probe. The laser source supplies a pair of 532 \textit{nm} wavelength (green) laser beams, separated horizontally by a distance of 60 \textit{mm} for velocity measurements in the streamwise direction (or \textit{x}-direction), and a pair of 561 \textit{nm} wavelength (yellow) laser beams, separated vertically by a distance of 60 \textit{mm} for velocity measurements in the wall normal  direction (radial, or \textit{y}-direction). The optical set-up allows for an ellipsoidal measurement volume 0.1\textit{mm} diameter and 0.3 \textit{mm} length with a 150 \textit{mm} focal length lens. A frequency shift of 80 \textit{MHz} is applied to each laser by a Bragg cell. The probe is connected by to the Burst Spectrum Analyser (BSA) signal processor for data acquisition with the Dantec BSA software. Polyamide seeding particles of 5 $\mu$\textit{m} average size manufactured by Dantec Dynamics are used for the experiments. The uncertainty of the LDA system measurements, according to the factory calibration certificate, is 0.1\% of the mean velocity being measured.
	
	We show the schematic of the LDA traverse in \hyperref[flowloop2]{Figure~\ref*{flowloop2}}. The main velocity measurements are performed along the duct centre plane, shown in \hyperref[flowloop2]{Figure~\ref*{flowloop2}} as measurement plane 1 (MP1). The measurement volume is set in a coincident position to MP1 by traversing the probe in the $z$-direction. The differences in refractive indices (RI) of the fluids, acrylic duct and air have to be considered to estimate the position of intersecting laser pairs in the fluid. We measured the refractive indices of all fluids with a Cole-Parmer Digital Refractometer, with a range from $n_r  = 1.3333$ to $1.5400$ RI. All polymer and surfactant fluids showed a RI of $n_r = 1.33$ which is the same as water. The position of the measurement volume in the fluid was then calculated with the refractive indices of the polymer and surfactant solutions and acrylic ($n_r =  1.49$), resulting in a measurement volume traverse of 0.69 \textit{mm} within the duct as the probe moves 0.50 \textit{mm} \citep{mitishita2021}. To further characterize the flows of polymers and surfactant solutions, we perform additional measurements in the measurement planes 2 and 3 (MP2 and MP3), which are located at $z/w  = 0.2$ and $z/w  = 0.6$ respectively. A full probe traverse is done from the bottom of the duct at $y/h = 0$ to the centre at $y/h = 1$. We only study fully turbulent flows in this paper, and thus we assume the velocity profile is symmetric across the duct centreline and no asymmetry effects of transitional flows in non-Newtonian fluids is found here, such as in \citep{escudier2009b, guzel2009a, jaafar2010}. The time-averaged statistics were also corrected for velocity bias via the transit-time weighted averaging method. Additional data on the verification of the experimental setup can be found in \citep{mitishita2021} and its supplementary material.
	
	\subsection{Fluid preparation} \label{s2.2}
	
	We outline the fluid preparation for the turbulent flow loop experiments. For more clarity, we separate the preparation protocol for the polymers and the surfactant solution.
	
	\subsubsection{Xanthan gum and partially hydrolyzed polyacrylamide} \label{s2.2.1}
	
	To formulate the drag-reducing polymer solutions, we use xanthan gum (XG), a rigid polymer supplied by CP Kelco, and partially-hydrolyzed polyacrylamide (HPAM) of commercial name Poly-plus RD, a flexible polymer supplied by MI-Swaco. We mixed both polymers in approximately 220L of tap water in the flow loop. The XG was prepared at a  concentration 2000 ppm (0.2\%w, or weight concentration in relation to the total weight of tap water in the tank), and the HPAM was prepared at a concentration of 500 ppm (0.05\%w). The tap water was recirculated at a high speed ($\approx$ 5 \textit{m/s}) while the correct amount of polymer powders were slowly poured in the tank. To ensure complete mixing after the addition of the polymers, we recirculated the solution for an additional 30 minutes at 3 \textit{m/s}. After mixing, the polymers were rested overnight prior to the experiments. Both of these polymers form clear solutions that are suitable for for laser doppler velocimetry measurements. The high concentrations of the polymer solutions were employed to guarantee maximum drag reduction and to reduce the effects of degradation due to re-circulation in the flow loop at high Reynolds numbers.
	
	\subsubsection{Viscoelastic surfactants} \label{s2.2.2}
	
	We employ the commercial gravel-packing surfactant J590, supplied by Schlumberger Oilfield Services in liquid form, of specific gravity listed as equal to water. The J590 contains primarily a mixture of the zwitterionic surfactant \textit{erucic amidopropyl dimethyl betaine} and \textit{isopropanol}. The rheological study of aqueous solutions with the J590 surfactant by \citet{gupta2021} confirms the similarities between the solutions used in this paper and EDAB solutions used in \citep{kumar2007, raghavan2012,beaumont2013}, in addition to micelle formation visualized by cryo-TEM. We use the J590 product as supplied, and it is mixed with tap water at 0.05, 1.00 and 1.35\%\textit{v} (volume percentage relative to 220\textit{l}) for flow loop experiments, to result in approximately 220L of surfactant solution. Note that the concentrations listed here refer to the J590 liquid, not of the surfactant EDAB itself, because the J590 product sheet does not reveal the exact surfactant concentration. We circulate the VES solution in the flow loop at a bulk velocity of approximately 6 \textit{m/s}, as surfactant requires high shear rates to completely mix with water. The final surfactant solution is slightly turbid after mixing. Given that our tap water is usually quite cool (below 20$^\circ$C), an additional step of recirculating while heating the fluid to approximately 30$^\circ$C over 8h was sufficient to make it more transparent for LDA experiments. After mixing, the surfactant solutions were rested for two days prior to each set of experiments. For consistency with the companion paper by \citet{gupta2021}, we refer our surfactant solutions as VES (viscoelastic surfactant) throughout this paper.
	
	\subsection{Rheometry} \label{s2.3}
	
	To support our flow loop studies, we characterize the rheology of all drag-reducing solutions with with a high-resolution Malvern Kinexus Ultra+ stress-controlled rheometer. We first outline the geometries used for the polymer solutions. We use a parallel plate geometry of 40 \textit{mm} in diameter with a 0.3 mm gap to allow for high shear rate measurements. The shear position in the parallel plates is $r/R = 0.75$, where $R$ is the radius. The fluid temperature is controlled by a Peltier system, with temperature tolerance of $\pm 0.1^{\circ}C$. Prior to experiments, the temperature of XG and HPAM is left to stabilize for 5 minutes in the parallel plates. We also performed small angle oscillatory shear (SAOS) experiments in the linear viscoelastic region, to characterize the elastic behaviour of the polymers in oscillatory shear flow. For the SAOS experiments, we use a C25 concentric cylinders geometry for increased torque measurement, with a cup of 27.5 \textit{mm} diameter and height of 62.50 \textit{mm}, and bob with 25 \textit{mm} diameter, 37.50 \textit{mm} height and cone angle of 15$^{\circ}$.  Due to the larger sample volume required for the concentric cylinders geometry ($\approx 18$ \textit{ml}), we waited 15 minutes for the temperature to stabilize.
	
	The VES solution exhibited edge instabilities during rheological tests with both cone-plate and parallel-plate geometries, so we employed a roughened (sandblasted) C34 concentric cylinders geometry to prevent edge effects and wall slip in all rheological experiments, with a cup of 37.00 \textit{mm} diameter and height of 66.00 \textit{mm}, and bob with 33.65 \textit{mm} diameter, 37.50 \textit{mm} height and cone angle of 15$^{\circ}$. The temperature of the VES is left to stabilize for 20 minutes in the concentric cylinders due to the large sample required ($\approx 36$ \textit{ml}). 
	
	\subsection{Turbulent flow experiments} \label{s2.4}
	
	Our experimental protocol consists of a comparison between water, 0.2\%\textit{w} xanthan gum, 0.05\%\textit{w} partially hydrolyzed polyacrylamide (hereafter named XG 0.2\% and HPAM 0.05\%), and the VES surfactant at 0.50, 1.00 and 1.35\%\textit{v} concentrations (hereafter named VES 0.50\%, VES 1.00\% and VES 1.35\%), all of which are semi-dilute. With all fluids listed, we measure the turbulent velocity profile (one traverse as specified at the end of \hyperref[s2.1]{Subsection~\ref*{s2.1}}) with LDA simultaneously with the pressure, flow rate and temperature. Before starting the velocity measurements, we circulate the fluid for approximately five minutes to ensure a steady-state. LDA measurements of velocity profiles are performed at bulk velocities of 2.8, 3.8 and 5.8 \textit{m/s} in the duct, with the exception of VES 0.50\%, which was investigated at 2.8 and 3.8 \textit{m/s}, but not at 5.8 \textit{m/s} due to the fact that TDR was very low already at 3.8 \textit{m/s}. We note that the bulk velocity parameters shown here are merely an approximation, as the actual measured bulk velocities differ slightly. Therefore, 14 sets of turbulent flow experiments were conducted for the present paper, with some of the water results being the same as presented in \citep{mitishita2021}. Due to the large volumes used in our loop, the same batch of each polymeric fluid is employed during measurements at all three bulk velocities listed here to minimize waste. However, we do present rheological experiments to quantify any degradation during the experiments. Micellar fluids at rest have the advantage of recovering their structure (and thus DR ability) after shearing \citep{gasljevic2007, gasljevic2017}. Therefore, each set of VES experiments was performed with the fluid initially at rest for at least 36 hours. 
	
	The mean wall shear stress is calculated by: 
	
	\begin{equation}  \label{eq-tauw}
		\tau_w = \frac{\Delta P D_h}{4 L},
	\end{equation}
	
	where $\Delta P$ is the pressure drop measurement and $L$ is the length between each pressure tap. The friction velocity is then calculated as $u_{\tau} = \sqrt{ \tau_w/ \rho }$. The wall shear stress is also used to calculate the percentage of drag reduction 
	
	\begin{equation}  \label{eq-DR}
		\%DR = \frac{ \tau_{w,w} - \tau_{w,DR} }{ \tau_{w,w} } \times 100,
	\end{equation}
	
	where $\tau_{w,w}$ is the mean wall shear stress of water and $\tau_{w,DR}$ is the mean wall shear stress of the drag-reducing fluid, both measured at the same bulk velocity $U_b$ measured by the flow meter. The density of all solutions is the same as water ($\rho$ = 999 \textit{kg/m$^3$}). With $U_b$ and the friction velocity we define the generalized Reynolds number and the frictional Reynolds number, respectively: 
	
	\begin{equation}  \label{eq-ReG}
		Re_G = \frac{\rho U_b D_h}{\eta_w},
	\end{equation}

	\begin{equation}  \label{eq-Retau}
		Re_\tau = \frac{\rho u_\tau h}{\eta_w},
	\end{equation}

	where $\eta_w = \tau_w/\dot{\gamma}_w$ is the viscosity of the fluid at the wall and $h$ is the duct half-height, which is also the boundary layer thickness in fully developed duct flow. The value of $\eta_w$ is used to define the Reynolds number in other studies of turbulent flows with non-Newtonian fluids \citep{escudier2009a,owolabi2017,singh2017b} where the viscosity of the fluid is spatially variant in a pipe or a duct. Note that $Re_G$ = $Re$ for the water experiments, in which case $\eta_w$ is constant. 
	
	With the polymer solutions, the mean shear rate at the wall $\dot{\gamma}_w$ can be obtained with the wall shear stress and steady flow curve data fit to a constitutive equation. The viscosity at the wall is also used to define the wall unit $y^+_0 = \eta_w /\rho u_{\tau}$, a viscous length scaled used to normalize the wall normal coordinate $y$, measured from the bottom wall of the duct. However, it is not straightforward to obtain $\eta_w$,  $y^+_0$ and $Re_G$ for turbulent flow of the VES solutions, due to the fact that the micellar structure appears to change significantly with the high shear rates in a turbulent duct flow. Thus, we provide a more detailed discussion of velocity profiles and additional rheological characterization of the VES in \hyperref[s4]{Section~\ref*{s4}}. The Reynolds numbers, friction velocity, fluid temperature, mean wall shear stress and \%DR values are presented in \hyperref[exp-t1]{Table~\ref*{exp-t1}} for water and the polymer solutions, where bulk velocities investigated in the loop are approximately the same for water and polymeric fluids: $U_b = 2.8,~3.8$ and $5.8~m/s$.
	
	\begin{table}
		\begin{center}
			\begin{tabular}{  c  c  c  c  c  c  c  }
				\hline
				U$_b$ [$m/s$] & $Re_G$ [-] & $u_\tau$ [$m/s$] & T [$^\circ$C]  &  $\tau_w$ [$Pa$] & $\eta_w$ [$Pa.s$]  & $DR$ [\%]  \\
				\hline
				\multicolumn{7}{c}{Water} \\			
				\hline
				2.86 & 105,480 & 0.134 & 23.6 & 17.9 & 0.0009 & - \\
				3.85 & 147,660 & 0.173 & 25.3 & 29.8 & 0.0009 & - \\
				5.86 & 232,810 & 0.249 & 26.8 & 61.7 & 0.0009 & - \\
				\hline
				\multicolumn{7}{c}{XG 0.2\%} \\	
				\hline
				2.79 & 30,256 & 0.085 & 26.4 & 7.27 & 0.0031 & 59 \\
				3.76 & 48,754 & 0.103 & 26.8 & 10.63 & 0.0026 & 64 \\
				5.72 & 87,983 & 0.139 & 27.6 & 19.37 & 0.0022 & 69 \\
				\hline
				\multicolumn{7}{c}{HPAM 0.05\%} \\	
				\hline
				2.81 & 40,729 & 0.081 & 24.7 & 6.59 & 0.0026 & 63 \\
				3.75 & 63,392 & 0.101 & 25.1 & 10.08 & 0.0023 & 66 \\
				5.67 & 112,449 & 0.133 & 25.9 &  17.75 & 0.0018 & 71 \\
				\hline
			\end{tabular} \caption{Measured bulk velocity, generalized Reynolds number, friction velocity, fluid temperature in the tank, wall shear stress and drag reduction percentage for water and polymer solutions.}
			\label{exp-t1}
		\end{center}
	\end{table}
	
	\section{Polymer and VES Rheology} \label{s3}
	
	\subsection{Polymer solutions} \label{s3.1}
	
\begin{figure}
		\centering
		\includegraphics*[width=80mm]{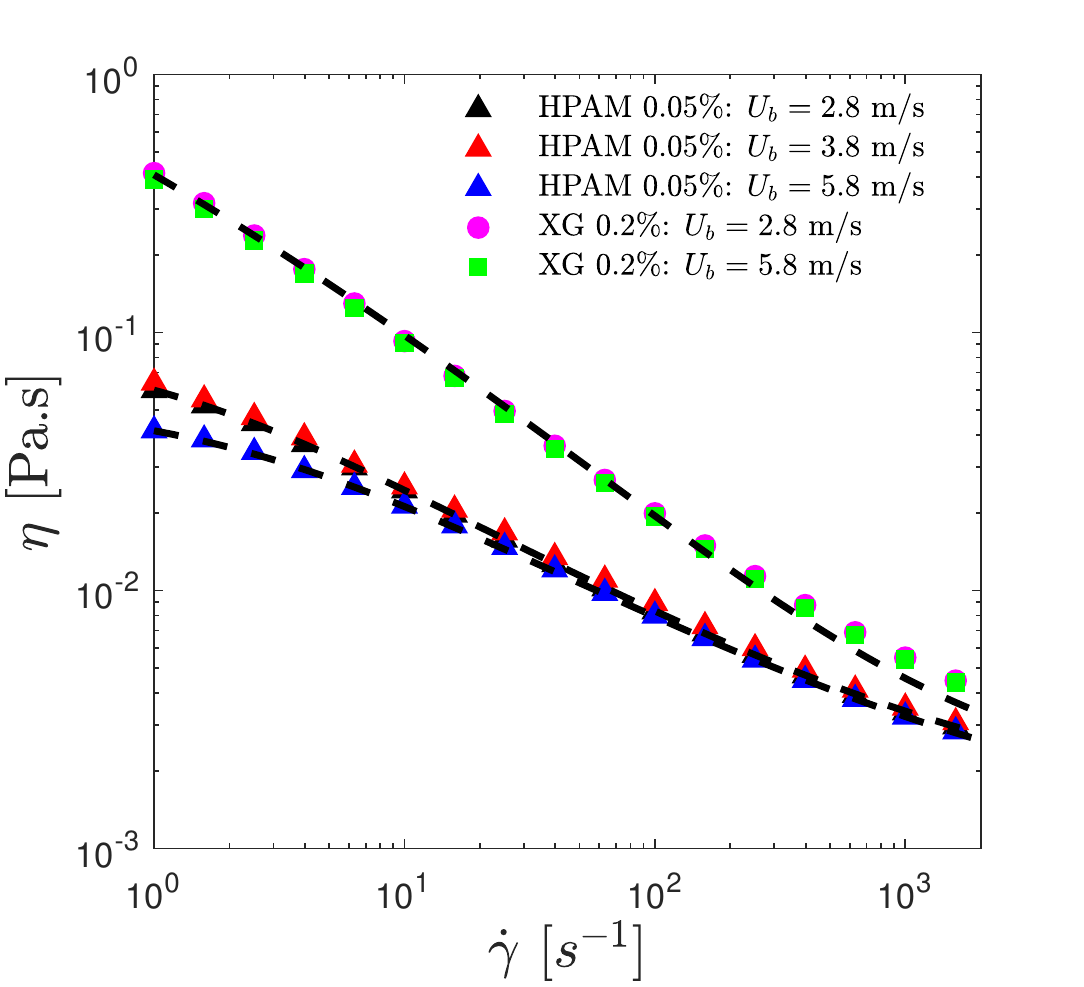}
		\caption{\fontsize{9}{9}\selectfont Steady flow curves for the 0.05\% HPAM and 0.2\% XG solutions. The samples were collected after flow loop experiments at the listed bulk velocities $U_b$. The dashed lines represent the curve fits to the Carreau-Yasuda equation.}
		\label{rheo-fig1}
	\end{figure}

	We characterize the polymer solutions via steady, shear-controlled flow curves and small amplitude oscillatory shear experiments. To quantify polymer degradation in the flow loop, each of the flow curves were measured with fluid samples taken after flow loop experiments at specified bulk velocities. Rheometry was performed at the average temperature in the flow loop tank. \hyperref[rheo-fig1]{Figure~\ref*{rheo-fig1}} shows the steady flow curves with all polymeric solutions employed in this study. Though shear viscosities of HPAM and XG solutions are quite different, they all show a shear-thinning behaviour under steady shear. The results suggest the XG solution is not significantly degraded when submitted to turbulent flow experiments at $U_b = 2.8,~3.8$ (not shown) and $5.8~m/s$. The HPAM solution appears to have degraded significantly only after a duct flow at $U_b = 5.8~m/s$. Considering these observations, the one Carreau-Yasuda curve was fitted for all xanthan gum flow curves. The CY model is written as:
	
	\begin{equation} \label{eq-CY}
		\eta = \eta_{\infty} + \frac{\eta_{0} - \eta_{\infty}}{(1 + (\lambda_{CY} \dot{\gamma})^a)^{n/a}},
	\end{equation}
	
	\noindent where $\eta_{0}$ is the zero-shear viscosity, $\eta_{\infty}$ is the infinite-shear-rate viscosity, $\lambda_{CY}$ is a constant with dimension of time, $n$ is a power-law index and $a$ is a fitting parameter \citep{yasuda1981}. Therefore, \hyperref[eq-CY]{Equation~\ref*{eq-CY}} is used to calculate $\dot\gamma_w$ based on the $\tau_w$ value, and thus $\eta_w$ with both XG 0.2\% and HPAM 0.05\% solutions to define the wall unit $y_0^+$ in \hyperref[s4]{Section~\ref*{s4.1}}.
	Due to the negligible degradation in the 0.05\% HPAM taken after the 2.8 and 3.8 \textit{m/s} experiments, a single CY curve could be fitted for both experiments. It has been well documented in the literature that rigid polymer solutions such as xanthan gum are quite resistant to mechanical scission due to the high shear stresses in turbulent pipe flows \citep{pereira2012,pereira2013,soares2015}, and a few studies relate apparent degradation of XG in experiments to destruction of molecular aggregates, a process named de-aggregation \citep{soares2019}. The degradation in the 0.05\% HPAM sample taken after a 5.8 \textit{m/s} experiment in the loop resulted in different fitting parameters for the CY equation. The CY fitting parameters are listed in \hyperref[rheo-t1]{Table~\ref*{rheo-t1}}.
	
	\begin{table}
		\begin{center}
			\begin{tabular}{  c  c  c  c  c  c }
				\hline
				U$_b$ [$m/s$] & $\eta_{0}$ [$Pa.s$] & $\eta_{\infty}$ [$Pa.s$] &  $\lambda_{CY}$ [$s$] & $a$ [-]  & $n$ [-]  \\
				\hline
				\multicolumn{6}{c}{XG - 2000ppm} \\	
				\hline
				2.8, 3.8, 5.8 & 9.425 & 0.0018 & 6.233 & 0.282 & 0.901 \\
				\hline
				\multicolumn{6}{c}{HPAM - 500ppm} \\	
				\hline
				2.8, 3.8 & 0.085 & 0.0014 & 0.999 & 1.039 & 0.540 \\
				5.8 & 0.054 & 0.0012 & 0.486 & 0.848 & 0.525 \\
				\hline
			\end{tabular} \caption{Carreau-Yasuda fitting parameters the viscosity of the XG solution at 2000 ppm and the HPAM solution at 500 ppm. The $U_b$ column shows during which LDA traverse in the pipe flow experiments the samples were collected.}
			\label{rheo-t1}
		\end{center}
	\end{table}

\begin{figure}
		\centering
		\includegraphics*[width=140mm]{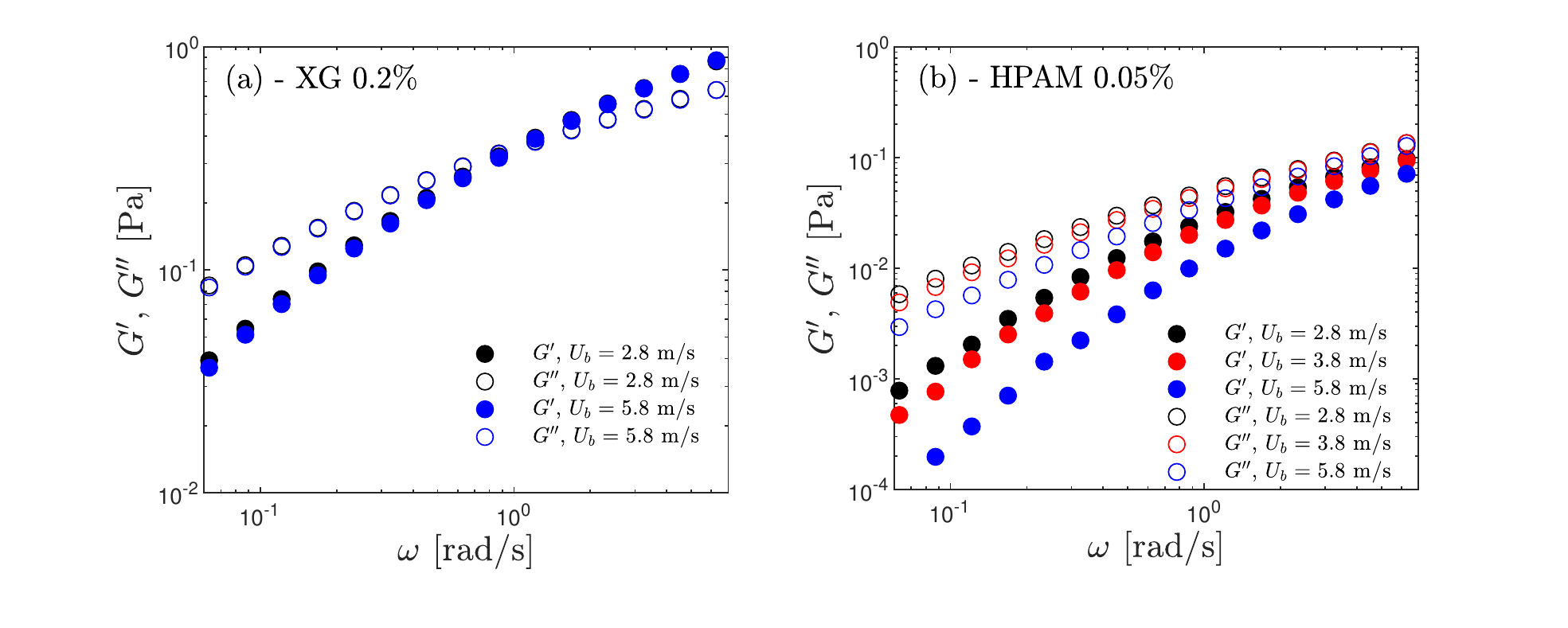} 
		\caption{\fontsize{9}{9}\selectfont Frequency sweeps in the linear viscoelastic regime with 0.05\% HPAM and 0.2\% XG solutions.}
		\label{rheo-fig2}
	\end{figure}

	The linear viscoelastic behaviour of the polymeric fluids was characterized via small-angle oscillatory shear (SAOS) experiments at different frequencies. The linear viscoelastic region was determined via stress-controlled amplitude sweeps with a frequency of 0.5 \textit{hz}, which resulted in SAOS amplitudes of $\gamma = 8\%$ for XG solutions and $\gamma = 10\%$ for HPAM solutions. The results from the SAOS experiments are presented in \hyperref[rheo-fig2]{Figure~\ref*{rheo-fig2}} as the storage and loss moduli - $G^{\prime}$ and $G^{\prime\prime}$ respectively - as a function of the oscillation frequency $\omega$. Similar to \hyperref[rheo-fig1]{Figure~\ref*{rheo-fig1}}, each sample studied with SAOS has been taken after sets of experiments at specific bulk velocities from the turbulent flow loop. We observe that both polymer solutions are viscoelastic with non-negligible values of $G^{\prime}$. The XG solution has dominant elastic behaviour ($G^{\prime} > G^{\prime\prime}$) for $\omega > 1~rad/s$ and dominant viscous behaviour ($G^{\prime} < G^{\prime\prime}$) at lower frequencies, and the curves for samples taken after flow loop experiments with $U_b = 2.8$ and $3.8~m/s$ are nearly identical, indicating negligible polymer chain scission in turbulent flow. From both the storage and loss moduli curves, we may define a characterisctic shear relaxation time $\lambda = 1/\omega$ at the cross over between the $G^{\prime}$ and $G^{\prime\prime}$ at $\omega \sim 1~rad/s$, resulting in $\lambda = 1~s$. Conversely, the HPAM solution has a dominant viscous behaviour in the linear viscoelastic regime in all frequencies investigated here. Due to inertia limitations, we are not able to measure the crossover point in the HPAM solution, which should be lower than the XG solution considering the maximum $\omega$ measured in our experiments, with $\lambda < 0.167~s$.
	
	\subsection{VES solutions} \label{s3.2}
	
\begin{figure}
		\centering
		\includegraphics*[width=150mm]{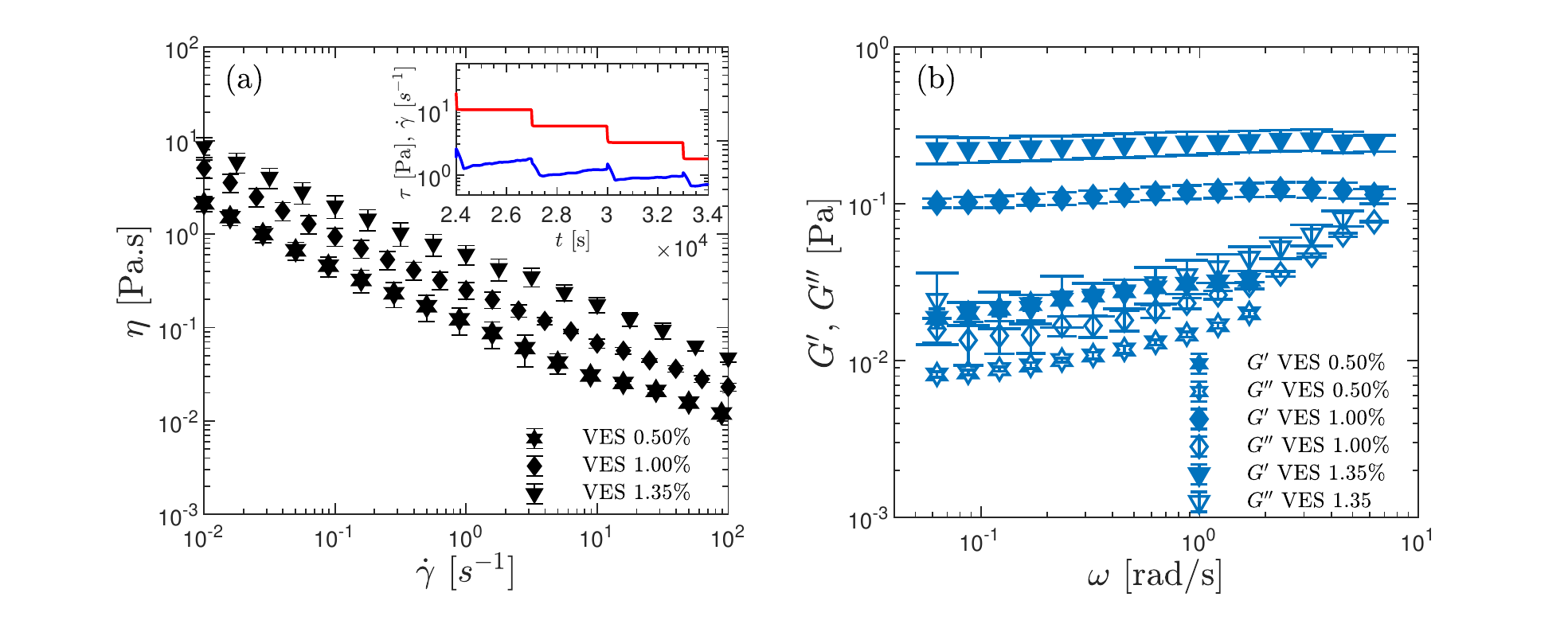}
		\caption{\fontsize{9}{9}\selectfont Steady flow curves (a) and frequency sweeps in the linear viscoelastic regime (b) with the 0.50, 1.00 and 1.35\% VES solutions. The inset in (a) shows shear stress (blue lines) and shear rate (red lines) measurements over time for one reproduction of the flow curve of the 1.35\% VES solution.}
		\label{rheo-fig3}
	\end{figure}

	The steady flow curves for the VES solutions are shown in \hyperref[rheo-fig3]{Figure~\ref*{rheo-fig3}} (a), for 0.05\%, 1.00\% and 1.35\% VES solutions. We imposed a pre-shear of $\dot\gamma = 500~s^{-1}$ for 5 minutes followed by a rest time of 5 minutes with a controlled stress of 0 \textit{Pa}. To obtain the flow curves, we imposed shear rates from $\dot\gamma = 500~s^{-1}$ to $0.01~s^{-1}$. To achieve a statistically steady flow, the shear stress was measured for 5 minutes at each imposed shear rate value, and the points were taken as the mean of the last 30 \textit{s} of shear stress measurements. The data over $100~s^{-1}$ appeared to be affected by inertial instabilities and is therefore not presented. Differently from the polymer solutions where a steady state was obtained in a few seconds with a constant imposed shear rate, the VES solution showed some instabilities in the stress response to the imposed shear rate, as seen in the inset of \hyperref[rheo-fig3]{Figure~\ref*{rheo-fig3}} (a), which are likely effects of elasticity. To account for the variations in the data due to these elastic instabilities, three flow curve experiments were performed and averaged as the final results in \hyperref[rheo-fig3]{Figure~\ref*{rheo-fig3}} (a). 
	
	The VES samples were selected regardless of the bulk velocity in flow loop experiments, but the temperatures probed in the rheometer were the same as the ones in the flow loop experiments at $U_b = 2.8~m/s$ (all VES flow loop experimental parameters are presented in a the next section).The error bars represent the standard deviation of all three flow curve experiments. Elastic instabilities during steady flows of wormlike micellar solutions were reported in \citet{beaumont2013} in an EDAB solution, and also in \citet{fardin2010} with a CTAB solution. \citet{beaumont2013} measured first normal stress differences $N_1$ in EDAB solutions, and observed large values with significant scatter in the data, which is evidence of elastic effects under shear. The mean values of viscosity as a function of shear rate show shear-thinning behaviour for all concentrations investigated. We note that, because of the relatively long stabilization time of 5 minutes for each point in the flow curve, we observe some time dependence in the measured shear stress over time, in agreement to what was presented in \citet{gupta2021} for the same material at high concentrations. Wormlike micelles are able to dynamically break and re-form its microstructure \citep{cates1990}, and we believe this time-dependence in the shear stress measurement may be a consequence of the micellar structure reconstruction under low shear rates. 
	
	We probe the viscoelastic behaviour of the three VES solutions in the linear regime with SAOS experiments. Prior to the SAOS experiments, we conduct a pre-shear of $\dot\gamma = 500~s^{-1}$ for 5 minutes followed by a rest time of 30 minutes at a controlled stress of 0 \textit{Pa}, following \citep{gupta2021}. The long rest times lead to a constant $G^{\prime}$ during experiments with a fixed frequency. The linear viscoelastic regime was determined via amplitude sweeps at a frequency of 0.5 \textit{hz}. The SAOS results under different imposed frequencies are presented in \hyperref[rheo-fig3]{Figure~\ref*{rheo-fig3}} (b). The results show a gel-like response to imposed oscillations in the linear viscoelastic regime, which is characterized by near constant values of the storage modulus $G^{\prime}$ as a function of frequency with an absence of a crossover point between $G^{\prime}$ and $G^{\prime\prime}$ in the frequency range investigated. The $G^{\prime\prime}$ values are also nearly a decade lower than $G^{\prime}$ and low phase angle $\delta$ values, akin to SAOS results in \citep{gutowski2012, giuseppe2015, mitishita2021} for Carbopol dispersions and \citep{raghavan2012,gupta2021} for an EDAB surfactant solution, which exhibits gel-like characteristics at room temperature. The $G^{\prime}$ and $G^{\prime\prime}$ data in conjunction with the VES flow curves hints to the presence of a yield stress, at least in the time-scale of our experiment. An apparent yield stress has been noted by \citep{kumar2007} in EDAB solutions, and in \citep{goyal2017} with a similar VES mixture to what we used in the current work. \citet{raghavan2012} conjectured that EDAB form stiff, long wormlike micelles which result in very long relaxation times in the linear viscoelastic regime, which is in contrast to a typical Maxwell viscoelastic fluid typical of most surfactant solutions, e.g.~\citep{haward2012,garcia2018}. \citet{gupta2021} also provided cryo-TEM images for the VES solutions (albeit in higher concentrations than the ones used in the present study), which show evidence of an entangled network of wormlike micelles longer than 1 $\mu m$, but without the cross links typically seen in Carbopol dispersions. The gel-like behaviour of $G^{\prime}$ and $G^{\prime\prime}$ are considered to be a consequence of these long and entangled micelles \citep{kumar2007}. We can estimate the relaxation time of a gel-like fluid in the linear viscoelastic regime from the following relation \citep{steller2016}:
	
	\begin{equation} \label{eq-lambda}
		\lambda = \lim_{\omega \to 0} \frac{ G^{\prime} }{\omega G^{\prime\prime} },
	\end{equation}
	
	\noindent where the values of $G^{\prime}$ and $G^{\prime\prime}$ as $\omega \to 0$ can be approximated to $G^{\prime}(\omega = 0.06~rad/s)$ and $G^{\prime\prime}(\omega = 0.06~rad/s)$ which is the lowest measured value of frequency in \hyperref[rheo-fig3]{Figure~\ref*{rheo-fig3}}, similarly to the analysis by \citet{oladosu2020}. From \hyperref[eq-lambda]{Equation~\ref*{eq-lambda}}, the relaxation times of the VES solutions in the linear viscoelastic regime are as follows: $\lambda  \sim 148~s$ for 1.35\% VES, $\lambda \sim 100~s$ for 1.00\% VES and $\lambda  \sim 38~s$ for 0.05\% VES. For reference, the relaxation time $\lambda$ for the VES 1.00\% is close to the value estimated by \citet{wang2017} for a 1.00\% EDAB solution. These relaxation times are substantially longer than $\lambda$ from both polymer solutions, and are evidence of the gel-like state in the linear viscoelastic regime. However, the gel-like rheology results and estimated relaxation times of the VES have to be interpreted with care, as it has been recently shown by \citet{gupta2021} that the EDAB/VES could be a highly viscoelastic fluid with very long relaxation times, and not a ``permanent" gel, with possible crossover of viscoelastic moduli at much lower frequencies. We acknowledge that our calculated $\lambda$ values for the VES solutions are limited by the fact that we cannot experimentally confirm if $G^{\prime}$ and $G^{\prime\prime}$ cross over at much lower $\omega$, which is certainly a possibility. However, given that both $G^{\prime}$ and $G^{\prime\prime}$ appear to become independent of frequency for $\omega < 0.1~rad/s$, we believe that \hyperref[eq-lambda]{Equation~\ref*{eq-lambda}} is adequate to estimate the relaxation time of the VES in the linear regime. Thus, we may be observing a gel-like rheology because of the relatively small timescales of the SAOS procedure in comparison to $\lambda$. In the next section, we present the turbulent flow studies with the characterized non-Newtonian solutions.
	
	\section{Velocity and Reynolds stress profiles of polymer and VES solutions}  \label{s4}
	
	\subsection{Velocity profiles and friction factor measurements} \label{s4.1}
	
	In this section we present the turbulence measurements of average velocities the LDA velocity measurements of XG 0.2\%, HPAM 0.05\% and VES at 0.05, 1.00 and 1.35\% concentrations and compare those to water (Newtonian) flows and DNS pipe flow data from \citet{ahn2015}, which is the same data for comparison as used in \citet{mitishita2021}. We decompose the instantaneous streamwise velocity measurements $U$ into a the sum of the mean streamwise velocity $\langle U \rangle$ and the turbulent velocity fluctuations $u$: $U = \langle U \rangle + u$ \citep{pope2001}. The same decomposition is applied to the wall normal velocity component: $V = \langle V \rangle + v$. 
	
\begin{figure}
		\centering
		\includegraphics*[width=80mm]{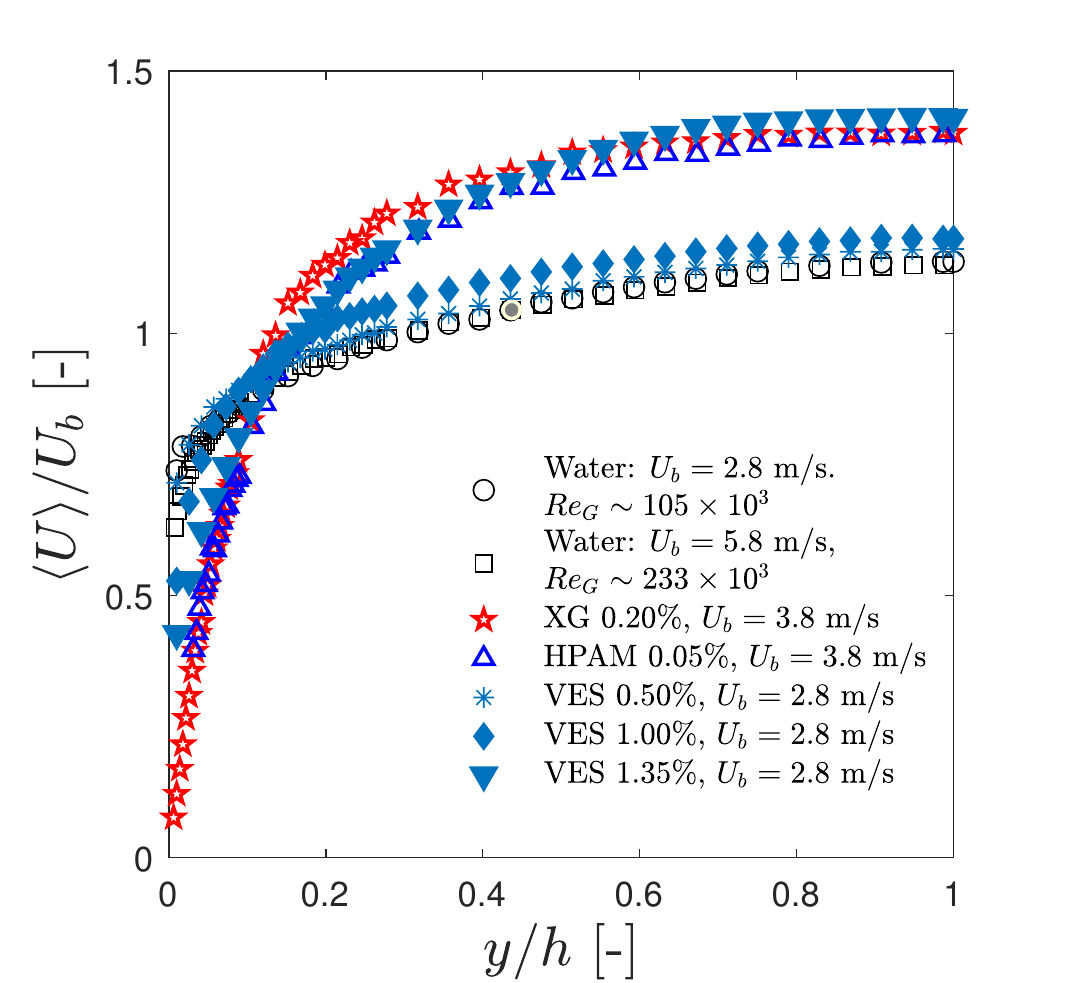}
		\caption{\fontsize{9}{9}\selectfont Velocity profiles of water, 0.05\% HPAM and 0.2\% XG, and 0.50, 1.00 and 1.35\% VES solutions normalized by $U_b$.}
		\label{stats-fig1}
	\end{figure}

	First, we present the mean velocity profiles for water at $U_b = 2.8$ and $3.8~m/s$, HPAM 0.05\% and XG 0.2\% at $U_b = 3.8~m/s$ and the VES 0.05, 1.00 and 1.35\% solutions at $U_b = 2.8~m/s$. The corresponding values of $Re_G$ for water and the polymers are listed in \hyperref[rheo-t1]{Table~\ref*{rheo-t1}} for reference. The parameters for VES solutions are presented later in this section, following a discussion on the near wall fluid rheology. From the results of \hyperref[stats-fig1]{Figure~\ref*{stats-fig1}}, both polymer solutions and also the VES 1.35\% achieve similar velocity profiles when normalized by the bulk velocity $U_b$ (which implies similar $\%DR$), while the water and VES 0.05\% mean velocity profiles are approximately the same. The VES 1.00\% velocities are somewhat higher than water. The velocity gradient for both polymers and the VES 1.35\% solution near the wall are noticeably lower than water, which is characteristic of a turbulent drag-reduced flow \citep{shaban2018}.
	
	It is not straightforward to compute the inner-scaled velocity profile $U^+ = \langle U \rangle/ u_{\tau}$ as a function of $y^+ = y/y^+_0$ with the VES results. If we apply the same approach used to obtain $\eta_w,~\dot\gamma$ and $Re_G$ for the HPAM and XG polymers and Carbopol solutions in \citep{mitishita2021} with the VES 1.35\% solution from the steady flow curve in \hyperref[rheo-fig3]{Figure~\ref*{rheo-fig3}}, and then compute $U^+$ and $y^+$, the resulting velocity profile does not match the XG and HPAM profiles, and is positioned to the left of Virk's asymptote in a $U^+$ \textit{vs.} $y^+$ plot (not shown). From the results of \hyperref[stats-fig1]{Figure~\ref*{stats-fig1}} where the velocities of both polymers and the VES 1.35\% solution are the same when normalized by $U_b$, we expect the inner-scaled velocity profiles to also match, assuming there are no significant errors in the pressure measurements, and thus $\tau_w$. The mismatch in the inner-scaled velocity profiles implies that the viscosity of the VES near the wall is not well represented by the steady flow curves, which results in an incorrect computation of the wall unit $y^+_0$. 
	
	We propose a few explanations for this error. First, we recognize that we are not able to reach steady shear rates above $\dot\gamma = 100~s^{-1}$ with the concentric cylinders geometry due to the appearance of inertial effects. We were able to measure shear rates of over $1000~s^{-1}$ with XG, HPAM, and also Carbopol solutions in \citep{mitishita2021} with the parallel plate geometry, but such measurements are not possible with the VES solutions because of edge fracture instabilities that occur at $\dot\gamma = O(1)$. Additionally, the laminar flow conditions in the rheometry experiments are very different from the fully turbulent flow conditions in the duct. One may then hypothesize that the micellar structure of the VES is also completely different in each experiment, since turbulent flows are well known to cause shear induced breakage in wormlike micelles \citep{gasljevic2007, li2005, wakimoto2018}, especially near the duct walls. Then, while the fluid in the rheometer is relatively structured due to the laminar conditions, the near-wall could be completely destructured, which explains why the $\eta_w$ computed from the steady flow curves is incompatible for defining $y^+_0$. Regarding the polymers studied here and the Carbopol in \citep{mitishita2021}, their molecular structure seems to be nearly unchanged in turbulent conditions due to the stronger molecular bonds than the wormlike micelles. Moreover, if the polymeric liquids are degraded in turbulent flow, the chain scission is permanent, so our calculation of $y^+_0$ with the steady flow $\eta_w$ is adequate for polymers.  
	
\begin{figure}
		\centering
		\includegraphics*[width=140mm]{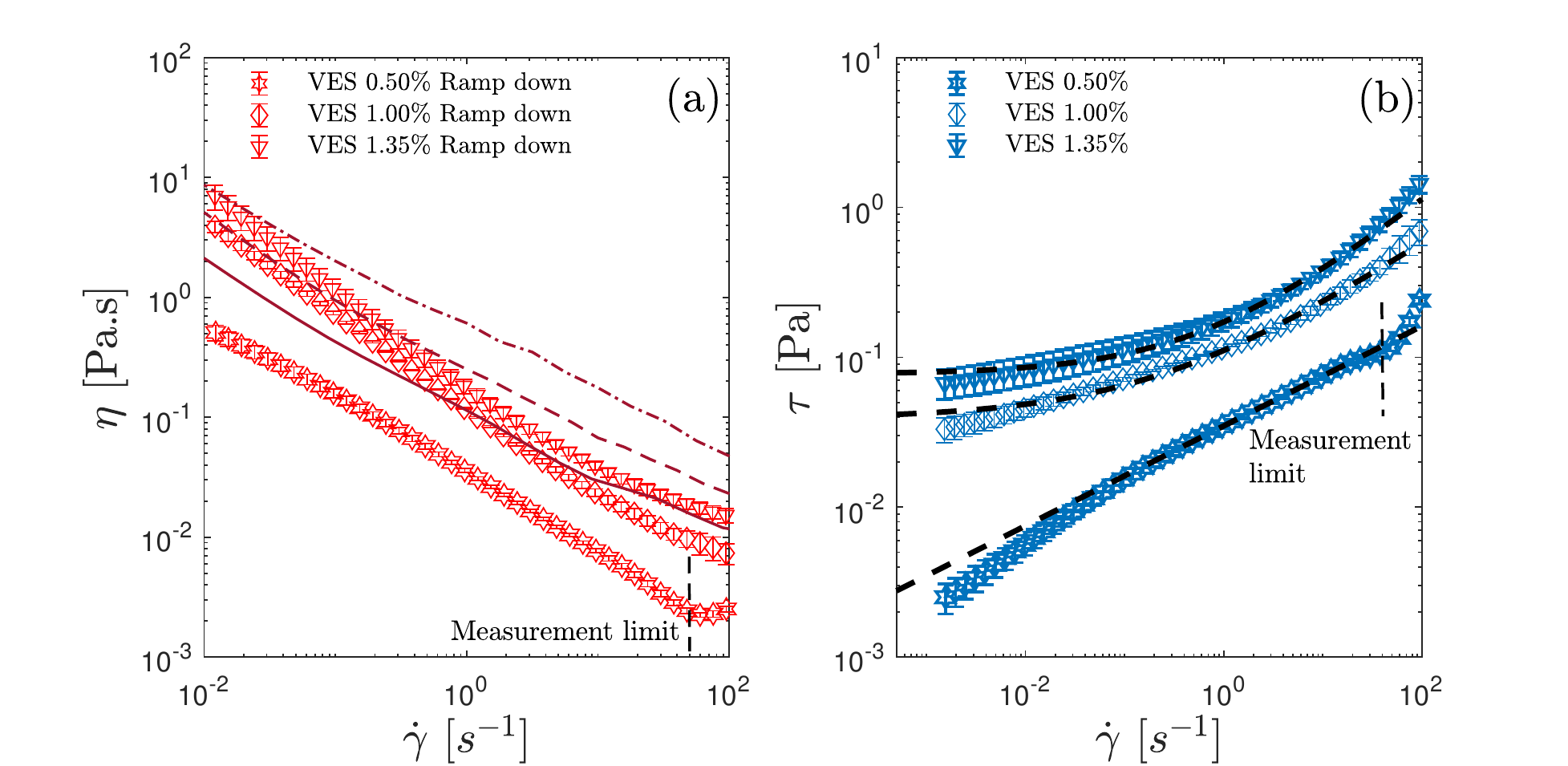}
		\caption{\fontsize{9}{9}\selectfont Shear-controlled ramp down flow curves for 0.50, 1.00 and 1.35\% VES solutions, with measurements of shear viscosity (a) and shear stress (b). In (a), the steady shear viscosities are shown by the full lines for VES 0.50\%, dashed lines for the VES 1.00\% and dash-dot lines for VES 1.35\%. In (b), the Herschel-Bulkley fits are represented by dashed lines. Measurement limits due to the appearance of inertial effects in the sample are shown by the vertical dotted lines.}
		\label{stats-fig2}
	\end{figure}

	To account for these difficulties with the VES, we estimate the rheological characteristics of the \textit{destructured} VES solutions near the wall using a distinct flow curve procedure from what was presented in \hyperref[s3.2]{Subsection~\ref*{s3.2}}, with the same concentric cylinders geometry. First, we impose a pre-shear at $\dot\gamma = 1000~s^{-1}$ for 5 minutes, to ensure the micellar structure in the sample is broken down. Next, \textit{without a prior rest period}, a ramp-down curve from $\dot\gamma = 100~s^{-1}$ to $0.001~s^{-1}$ for 3 minutes. The results are presented in \hyperref[stats-fig2]{Figure~\ref*{stats-fig2}}, and suggest the presence of a yield stress of similar magnitude than in \hyperref[rheo-fig3]{Figure~\ref*{rheo-fig3}}, but viscosities at high shear rates are much lower. We cannot consider the resulting flow curve of \hyperref[stats-fig2]{Figure~\ref*{stats-fig2}} as representative of a steady state measurement, but simply as an approximation of the rheology of the near-wall micellar fluid, because this procedure does not allow time for structure re-formation during the experiment. Therefore, we avoid the time-dependency that happens during lengthy steady shear flow curves, such as the shear stress measurements in the inset of \hyperref[rheo-fig3]{Figure~\ref*{rheo-fig3}} and also the results from the steady flow curves from \citep{gupta2021} at low shear rates.
	
	We can define $\eta_w,~\dot\gamma$ and $Re_G$ for the VES solutions by fitting the Herschel-Bulkley model to the flow curves of VES 1.35\% and VES 1.00\%, and the Power-Law model to the VES 0.50\% results of \hyperref[stats-fig2]{Figure~\ref*{stats-fig2}} (b). With the $\tau_w$ results from the flow loop experiments, we can compute $\dot\gamma_w$ and then $\eta_w$. Finally, we calculate the values of $Re_G$ and $y^+_0$ which were defined previously in \hyperref[s2.4]{Subsection~\ref*{s2.4}}. The Power-Law the Herschel-Bulkley models are represented by the following equations:
	
	\begin{equation} \label{eq-PL}
		\tau = K \dot\gamma^{n}, 
	\end{equation}
	
	\begin{equation} \label{eq-HB}
		\tau = \tau_y + K \dot\gamma^{n},\text{if}~\tau > \tau_y 
	\end{equation}
	
	\noindent where $\tau_y$ is the yield stress, $K$ is the consistency index and $n$ is the power law index. Note that in \hyperref[eq-HB]{equation~\ref*{eq-HB}}, $\dot\gamma = 0$ if $\tau \leq \tau_y$. The fitting parameters for \hyperref[eq-PL]{Equation~\ref*{eq-PL}} and \hyperref[eq-HB]{Equation~\ref*{eq-HB}}, considering the ramp-down flow curves for the VES fluids in  \hyperref[stats-fig2]{Figure~\ref*{stats-fig2}}, are presented in \hyperref[stats-t1]{Table~\ref*{stats-t1}}.
	
	\begin{table}
		\begin{center}
			\begin{tabular}{  c  c  c  c  }
				\hline
				Concentration & $\tau_y$ [$Pa$] & $K$ [$Pa.s^n$] &  $n$ [-]\\
				\hline
				0.50\% & - & 0.035 & 0.333 \\
				1.00\%  & 0.039 & 0.072 & 0.438 \\
				1.35\%  & 0.077 & 0.095 & 0.523 \\
				\hline
				
			\end{tabular} \caption{Power-law and Herschel-Bulkley equation parameters for the ramp-down flow curves for 0.50, 1.00 and 1.35\% VES solutions.}
			\label{stats-t1}
		\end{center}
	\end{table}
	
	The values of $Re_G$, bulk velocity, friction velocity, fluid temperature, mean wall shear stress, the viscosity at the wall and percentage of drag reduction are presented in \hyperref[stats-t2]{Table~\ref*{stats-t2}} for all three VES concentrations, for turbulent flow loop experiments at the listed bulk velocities. We also remind the reader that the same experimental quantities for turbulent flows of polymer solutions are listed in \hyperref[exp-t1]{Table~\ref*{exp-t1}}. Note that our estimations revealed $\eta_w$ values equal to water for all VES 0.50\% and 1.00\%, and the VES 1.35\% only at $U_b = 5.8~m/s$. With these experiments, we believe that the micelles near the wall are completely broken down, and the near-wall fluid is essentially a solvent/water layer due to the very high average shear rate at the wall. This explanation is plausible because micellar aggregation is formed by weak interactions that are easily broken by shear and extensional forces, whereas polymer molecules are formed by strong covalent bonds \citep{rothstein2008}.
	
	The profiles of the local time-averaged streamwise velocities of water, polymer solutions and VES solutions, normalized with the friction velocity ($U^+ = \langle U \rangle/ u_{\tau}$) are plotted against the wall normal position normalized by the wall unit ($y^+ = y/y^+_0$). The error bar in the water velocity profile at $U_b = 5.8~m/s$ represents the variation of $\tau_w$ across the rectangular duct, considering the fact that it is variant along its perimeter, unlike a pipe where $\tau_w$ is uniform. More details about the discussion on the effects of $\tau_w$ in our duct can be found in \citep{mitishita2021}.
	
	\begin{figure}
		\centering
		\includegraphics*[width=100mm]{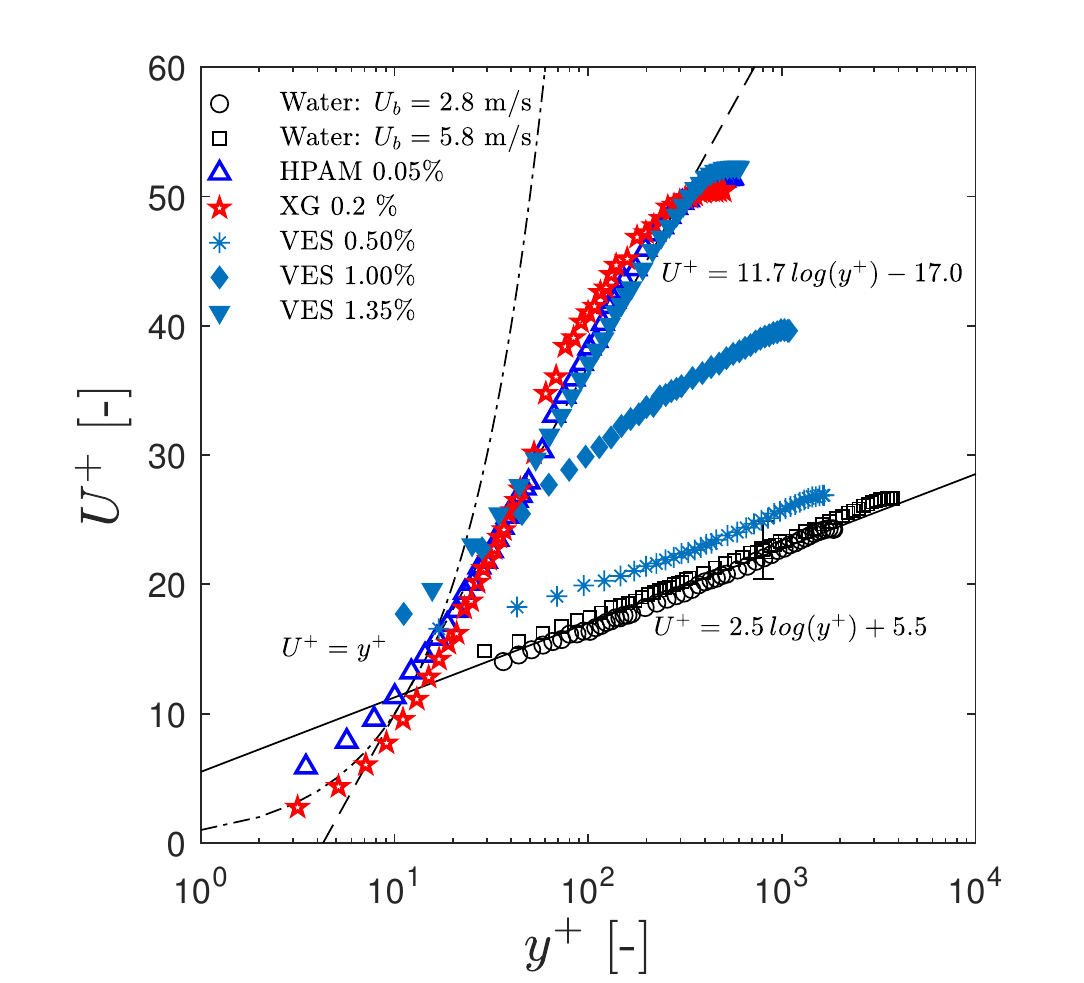}
		\caption{\fontsize{9}{9}\selectfont Velocity profiles of water, 0.05\% HPAM and 0.2\% XG, and 0.50, 1.00 and 1.35\% VES solutions normalized in wall units.}
		\label{stats-fig3}
	\end{figure}

	\begin{table}
		\begin{center}
			\begin{tabular}{  c  c  c  c  c  c  c  }
				\hline
				U$_b$ [$m/s$] & $Re_G$ [-] & $u_\tau$ [$m/s$] & T [$^\circ$C] & $\tau_w$ [$Pa$] & $\eta_w$ [$Pa.s$]  & $DR$ [\%] \\
				\hline
				\multicolumn{7}{c}{VES 0.50\%} \\			
				\hline
				2.86 & 101,995 & 0.1146 & 23.6 & 13.12 & $\sim$0.0009 & 26 \\
				3.85 & 140,634 & 0.1571 & 24.1 & 24.65 & $\sim$0.0009 & 17 \\
				\hline
				\multicolumn{7}{c}{VES 1.00\%} \\	
				\hline
				2.76 & 97,517 & 0.0823 & 21.8 & 6.77 & $\sim$0.0009 & 62 \\
				3.77 & 134,131 & 0.1497 & 22.1 & 22.38 & $\sim$0.0009 & 25 \\
				5.77 & 212,124 & 0.2238 & 23.4 & 50.02  & $\sim$0.0009 & 19 \\
				\hline
				\multicolumn{7}{c}{VES 1.35\%} \\	
				\hline
				2.75 & 59,420 & 0.0737 & 21.9 & 5.52 & 0.0016 & 69 \\
				3.75 & 137,459 & 0.0743 & 23.4 & 13.01 & 0.0011 & 56 \\
				5.75 & 212,703 & 0.2044 & 23.7 & 41.73 & $\sim$0.0009 & 32 \\
				\hline
				
			\end{tabular} \caption{Measured bulk velocity, generalized Reynolds number, friction velocity, temperature, pressure drop, wall shear stress, fluid viscosity at the wall, and drag reduction percentage (in this order) from flow loop experiments for all three VES solutions.}
			\label{stats-t2}
		\end{center}
	\end{table}
	
	We start our analysis of the inner-scaled velocity profile by comparing the velocity measurements from HPAM 0.05\%, XG 0.02\% ($U_b = 3.8~m/s$) and VES 1.35\% ($U_b = 2.8~m/s$), which according to our measurements, have $Re_G$ values of the same order of magnitude. As hinted by the outer-scaled velocity measurements of \hyperref[stats-fig1]{Figure~\ref*{stats-fig1}} as well as the \%DR measurements, the velocity profiles of HPAM 0.05\%, XG 0.02\% and VES 1.35\% reach Virk's asymptotic profile for maximum drag reduction (MDR). The collapse of all three curves appears to be independent of both inner or outer scaling, and suggests that the our different experimental protocol for the flow curve of the VES solutions was an adequate approach. The VES 1.00\% velocity profile is in the high drag reduction (HDR, \%DR $> 40$) regime, with a reasonably higher $U^+$ profile than water. The velocity profile for the VES 0.50\% hint to low values of drag reduction (LDR, \%DR $< 40$), where the velocity profiles approaches a logarithmic line similar to water, but with slightly higher values. We can evaluate the drag reduction in each experiment set by calculating the Fanning friction factor:
	
	\begin{equation} \label{eq-f}
		f = \frac{2 \tau_w}{\rho U_b^2}.
	\end{equation}
	
\begin{figure}
		\centering
		\includegraphics*[width=120mm]{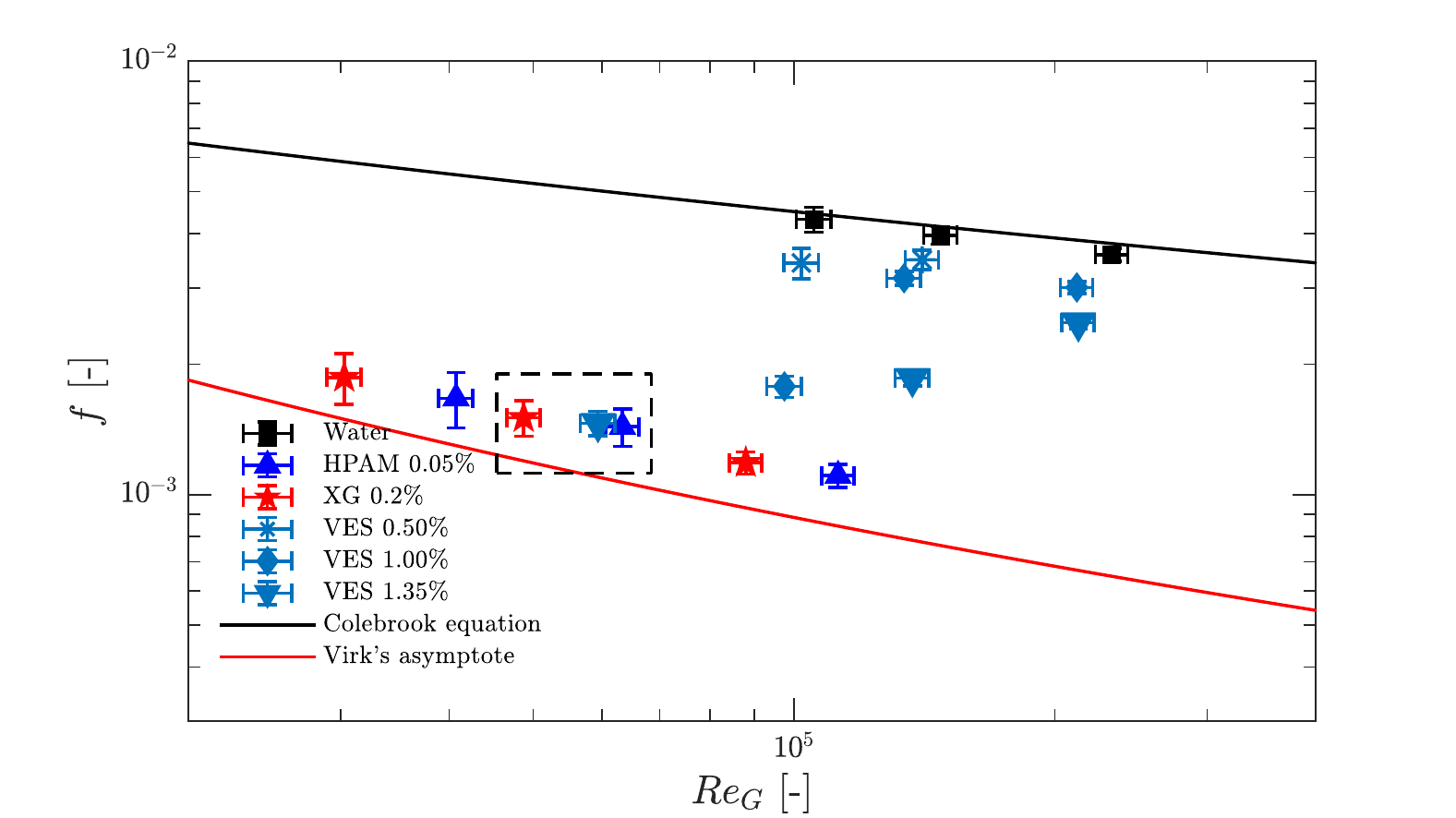}
		\caption{\fontsize{9}{9}\selectfont Friction factor as a function of the generalized Reynolds number for 0.05\% HPAM and 0.2\% XG solutions, and 0.50, 1.00 and 1.35\% VES solutions. The dashed rectangle corresponds to the MDR flows at which we compare the flow of surfactant and polymer solutions. The error bars represent the estimated propagation of measurement uncertainties \citep{kline1953} from the pressure transducers, flow meter and cutting tolerances from the acrylic duct (1 \textit{mm}) and the uncertainty from the viscosity measurements of the rheometer, of approximately $\pm 3\%$.}
		\label{stats-fig4}
	\end{figure}

	We plot the friction factor measurements of all fluids investigated against the generalized Reynolds number $Re_G$ in \hyperref[stats-fig4]{Figure~\ref*{stats-fig4}}. The water results are the ones presented in \citep{mitishita2021}. As a reference of Newtonian fluids, we also show the Colebrook line for turbulent flow friction factors, calculated based on the hydraulic diameter $D_h$ of our duct:
	
	\begin{equation} \label{eq-colebrook}
		\frac{1}{ \sqrt{f} }  = -4.0 \, log \, \left( \frac{\epsilon/D_h}{3.7} + \frac{ 1.255 }{ Re_G \sqrt{f} } \right),
	\end{equation}
	
	\noindent where the roughness $\epsilon$ is neglected for our acrylic duct. We also present Virk's line as a reference for the friction factor of polymer solutions at MDR, which represents an empirical approximation to the upper bound for polymer drag reduction:
	
	\begin{equation} \label{eq-virk}
		\frac{1}{ \sqrt{f} }  = 19.0 \, log \, \left( Re_G \, \sqrt{f} \right)  - 32.4.
	\end{equation}
	
	In \hyperref[stats-fig4]{Figure~\ref*{stats-fig4}}, the friction factor measurements reveal that the HPAM 0.05\% and XG 0.2\% solutions reach MDR under all conditions investigated in the flow loop. This was expected from XG 0.2\% considering the fact that the rheology results do not indicate degradation after flow loop experiments, likely due to loss of polymer aggregates instead of chain scission as indicated by \citep{soares2015}. From the HPAM 0.05\% results, the amount of fluid degradation observed in the rheology experiments did not incur in less drag reduction when $U_b$ is increased to $5.8~m/s$ ($Re_G \sim 112 \times 10^3$). This implies that even the degraded HPAM 0.05\% was sufficiently elastic to reach MDR under turbulent flow at $Re_G \sim 112 \times 10^3$. The results in the dashed-line rectangle represents HPAM 0.05\%, XG 0.2\% and VES 1.35\% fluids at MDR in the same conditions as shown in \hyperref[stats-fig3]{Figure~\ref*{stats-fig3}}. with friction factor and Reynolds number values close to each other. Therefore, the comparison of other turbulent quantities such as Reynolds stresses between the HPAM, XG and VES in these conditions is reasonable. 
	
	The micellar fluids behave quite differently as $Re_G$ increases. The VES 0.50\% exhibits very low levels of DR, with \%DR = 26 at $Re_G \sim 101 \times 10^3$ and \%DR = 17 at $Re_G \sim 140 \times 10^3$. We therefore expect breakdown of the gel-like micellar structure and subsequent loss of viscoelasticity in all VES 0.50\% experiments. The VES 1.00\% better resists breakage at higher $Re_G$, as observed from the high DR of 62\% at $Re_G \sim 97 \times 10^3$, but its drag reduction ability greatly decreases at ($Re_G \sim 134 \times 10^3$) or above that. As we increase the concentration of VES to 1.35\%, the MDR asymptote is reached at a similar $Re_G$ to the other two polymer solutions, as indicated by the rectangle in \hyperref[stats-fig4]{Figure~\ref*{stats-fig4}}, at $Re_G \sim 59 \times 10^3$. However, the VES 1.35\% is significantly degraded during the flow loop experiments at $Re_G \sim 137 \times 10^3$ ($\%DR = 56$) and $Re_G \sim 212 \times 10^3$ ($\%DR = 32$). Specifically at $Re_G \sim 212 \times 10^3$, the loss of drag reduction ability means that the \%DR of the VES 1.35\% solution is comparable to the VES 0.50\% case at $Re_G \sim 101 \times 10^3$. 
	
	From an industrial perspective, these results give important information about the advantages and drawbacks of the EDAB/VES micellar fluid at the concentrations investigated here. The main advantage, as mentioned previously, is the ability of the micelles to re-form its entangled structure after scission in turbulent flows. Evidence of this characteristic in the results of present study comes from the fact that the samples used in the rheological measurements were used at random, and collected at different flow velocities in the flow loop. Repeated flow curves from \hyperref[rheo-fig3]{Figure~\ref*{rheo-fig3}} do not indicate permanent degradation, and therefore the VES fluids can be successfully re-used in recirculating flow loop applications without permanent loss of DR ability. However, the HPAM 0.05\% and XG 0.2\% are more resistant to degradation, as they are able to maintain maximum drag reduction at high Reynolds numbers when compared to VES. So even though the polymer solutions degrade permanently due to chain scission, they appear to be more effective resisting a loss of \%DR due to an increase in Reynolds number in a turbulent flow, at least in the semi-dilute concentrations investigated here. Increasing the surfactant volume concentration in a VES solution seems to improve resistance to micelle breakage, as seen in experiments with different surfactants such as \citep{tuan2013b,tamano2015, hara2017}, and in our results from \hyperref[stats-fig4]{Figure~\ref*{stats-fig4}}. We have not investigated whether or not the resistance to breakage due to increased $Re_G$ values in our VES is affected by other changes such as temperature or addition of salts and a co-surfactant (a common practice in gravel-packing operations according to \citet{goyal2017}), so further study of the influence of these paremeters is required. For the next sections of the paper, we analyze the velocity and Reynolds stress profiles of the VES solutions and also compare the VES 1.35\% at $U_b = 2.8~m/s$ with the polymer fluids near MDR, at similar Reynolds numbers.
	
	\subsection{Reynolds stresses} \label{s4.2}
	
\begin{figure}
		\centering
		\includegraphics*[width=140mm]{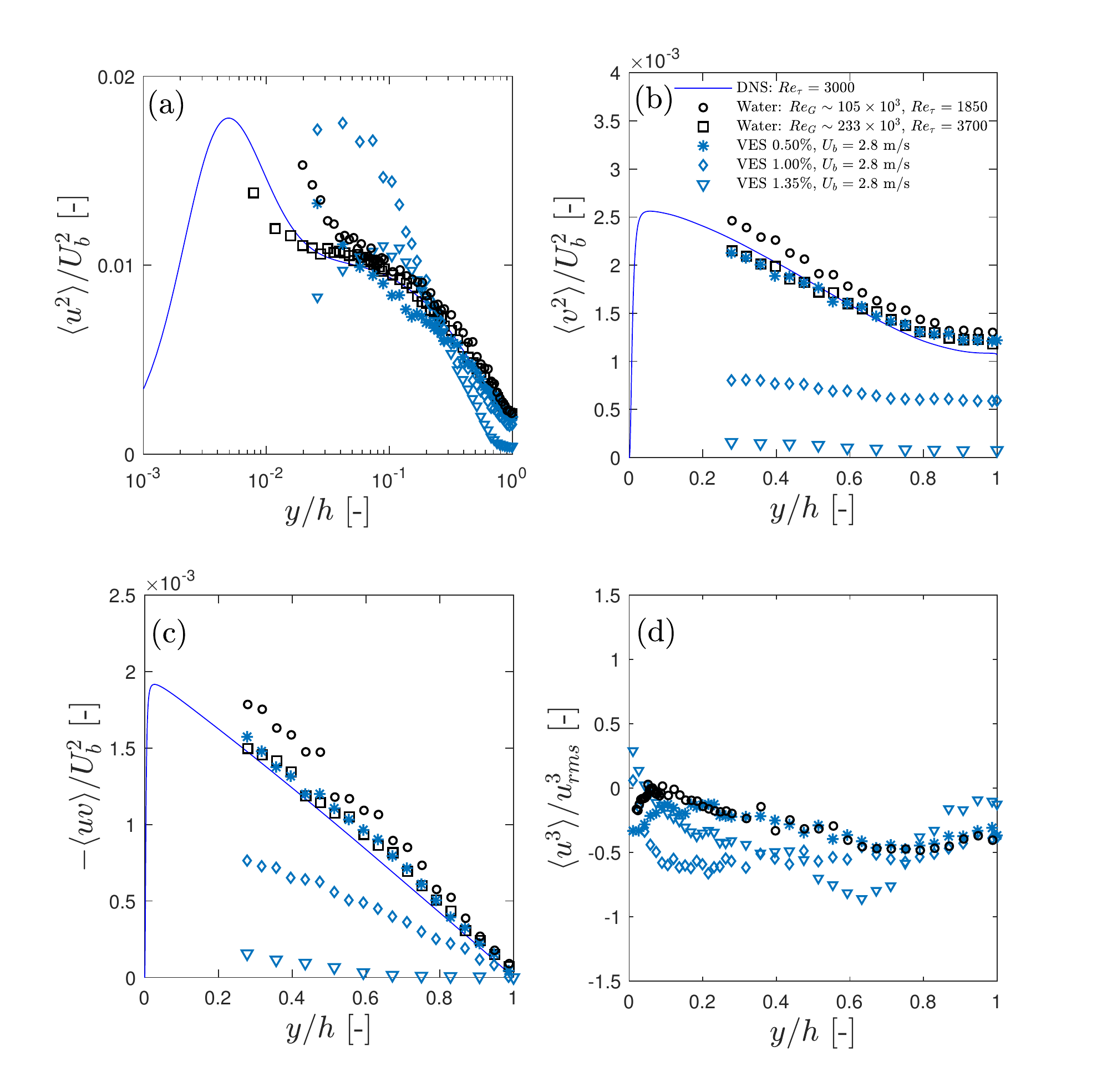}
		\caption{\fontsize{9}{9}\selectfont Streamwise (a), wall normal (b), shear Reynolds stresses and skewness (d) of streamwise fluctuations of water and 0.50, 1.00 and 1.35\% VES solutions.}
		\label{stats-fig5}
	\end{figure}

	We present measurements of Reynolds stresses to further clarify the effect of increasing the concentration of VES in a turbulent duct flow. Data from turbulent flows of HPAM 0.05\% and XG 0.2\% solutions are also presented, and serve both as a comparative study and to aid interpretation of our results. We begin our analysis with measurements of streamwise Reynolds stresses, $\langle u^2 \rangle$, wall-normal Reynolds stresses, $\langle v^2 \rangle$, Reynolds shear stresses, $\langle -uv \rangle$, and skewness $\langle u^3 \rangle/\sqrt{u^2}^3 = \langle u^3 \rangle/u_{rms}^3 $ of velocity fluctuations of all three VES concentrations in \hyperref[stats-fig5]{Figure~\ref*{stats-fig5}}. All turbulent quantities shown are normalized by the bulk velocity $U_b$ except for the skewness. The friction velocity $u_{\tau}$ is not used for normalization of turbulence data due to the fact that the wall shear stress is not homogeneous across the duct cross section, and a more detailed discussion about this matter is found in \citep{mitishita2021}. To complement the VES data, water measurements are presented alongside pipe flow DNS data from \citep{ahn2015} at $Re_{\tau} = 3000$ as a way of benchmarking our own experimental data. Additional verification of the experimental setup can be found in the supplementary material of \citep{mitishita2021}. 
	
	The results in \hyperref[stats-fig5]{Figure~\ref*{stats-fig5}} (a) show very similar $\langle u^2 \rangle$ profiles of VES 0.50\% and water at $U_b = 2.8~m/s$ at approximately the same Reynolds numbers. This support our claim that the micelles in the VES 0.50\% are likely degraded, with rheology approximately the same as water in a turbulent flow. It is important to note that we do not have a high enough spatial resolution to measure the peak in $\langle u^2 \rangle$ in water and VES 0.05\%, being so close to the wall. Increasing the concentration to 1.00\% results in a peak of $\langle u^2 \rangle$ close to the water DNS results, but seen farther away from the wall. The appearance of the $\langle u^2 \rangle$ peak farther from the wall when compared to Newtonian fluids is well documented in polymer drag reduction flows, especially at high DR (\%DR $> 40$), such as \citep{escudier2009a, mohammadtabar2017} and surfactants \citep{warholic1999}. At MDR, the VES 1.35\% shows a decreased peak in $\langle u^2 \rangle$ farther away from the wall compared to both Newtonian and VES 1.00\%. The presence of a $\langle u^2 \rangle$ away from the wall is a consequence of a thickened buffer layer at MDR, as reported by numerous investigations of polymer MDR such as \citep{escudier2009a,graham2014, shaban2018}. Additionally, the $\langle u^2 \rangle$ values decrease quite significantly across the entire duct at $y/h > 0.3$, because of the reduced turbulence and weakened vortical structures in the turbulent core \citep{xi2010,tamano2010,tuan2017} at MDR. 
	
	The effect of VES concentration is quite noticeable in the $\langle v^2 \rangle$ profiles shown in \hyperref[stats-fig5]{Figure~\ref*{stats-fig5}}(b), where $\langle v^2 \rangle$ of the VES 1.35\% are very close to zero, at least in the measurable duct positions with the LDA setup, which is characteristic of the MDR state \citep{warholic1999}. A decrease in concentration lowers the \%DR and consequently results in an increase in the $\langle v^2 \rangle$ values. For instance, at \%DR = 26 from the VES 0.50\% results, the measured wall-normal Reynolds stresses are slightly lower than the water results. The $-\langle uv \rangle$ results in \hyperref[stats-fig5]{Figure~\ref*{stats-fig5}} (c) follow the same trend as $\langle v^2 \rangle$, where the Reynolds shear stresses are very low at MDR with the VES 1.35\%, and increase with the reduction of VES concentration. The skewness results in \hyperref[stats-fig5]{Figure~\ref*{stats-fig5}} (d) are quite interesting in the sense that the VES 0.05\% results are nearly the same as water except in the near-wall region. Skewness of $u$ change significantly once the VES concentration is increased to 1.00\%, where $\langle u^3 \rangle/u_{rms}^3 \sim 0.5$ from $y/h \sim 0.1$ to $0.8$. Near the turbulent core, the skewness of VES 1.00\% becomes the same as water. The skewness profile again changes with the VES at 1.35\% concentration, where one sees a peak in positive skewness near the wall. This observation could correlate to increased sweep events ($u > 0, v < 0$) near the wall at $y/h < 0.1$, characterized by high speed fluid moving towards the wall \citep{wallace1972}. However, we are not able to simultaneously measure $u$ and $v$ at $y/h < 0.3$ with our LDA setup, and we can only speculate whether or not sweep events are enhanced. Near the centreline at $y/h = 1$, the skewness of $u$ is close to zero at MDR, which means that the distribution of velocity fluctuations is not skewed to neither positive or negative values. Together with the $\langle u^2 \rangle$ and $\langle v^2 \rangle$, the skewness at MDR suggests a large decrease in turbulent motions in the centre of the duct. 
	
	\begin{figure}
		\centering
		\includegraphics*[width=140mm]{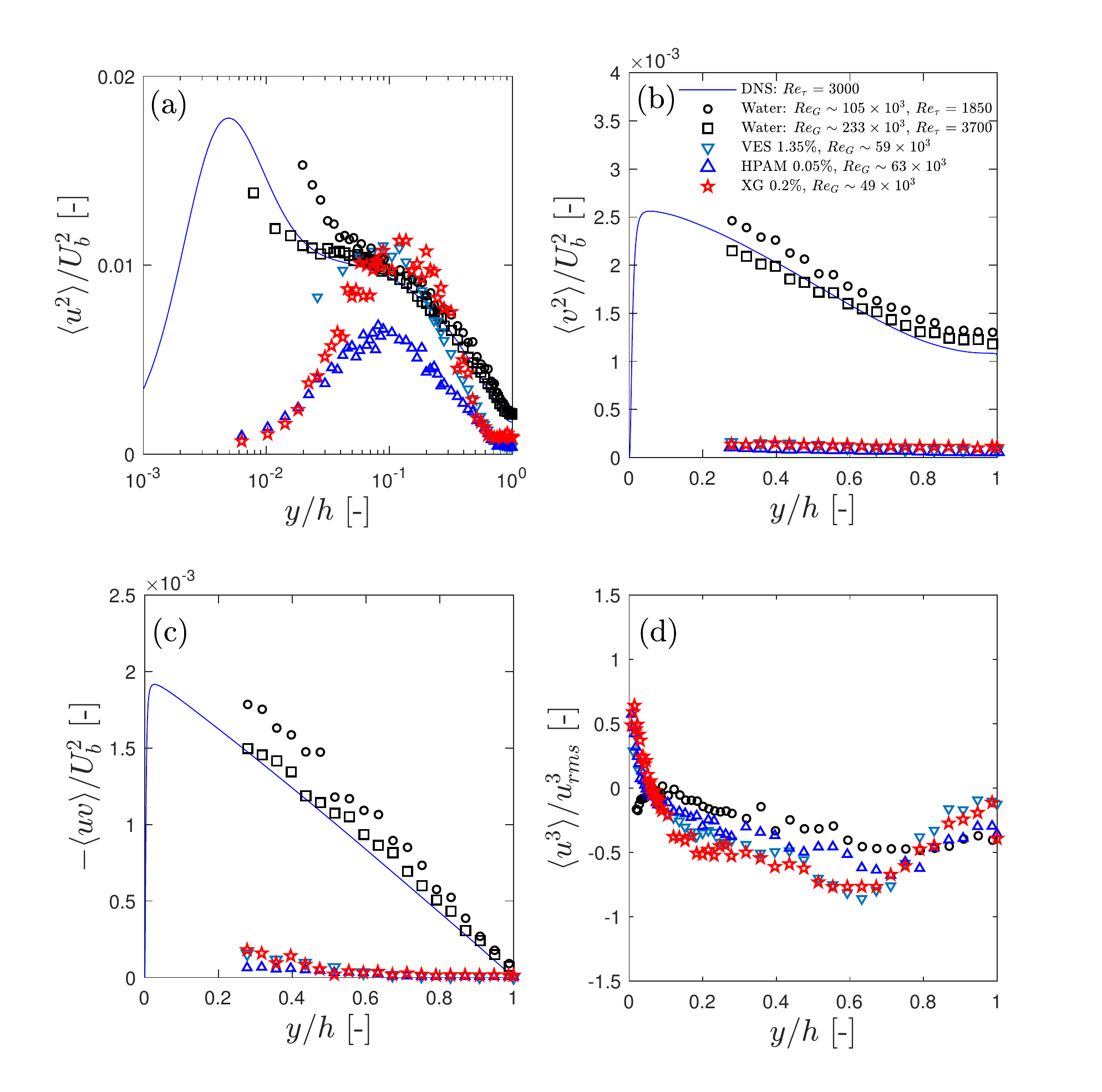}
		\caption{\fontsize{9}{9}\selectfont Streamwise (a), wall normal (b), shear Reynolds stresses (c) and skewness (d) of streamwise fluctuations of 0.05\% HPAM, 0.2\% XG and 1.35\% VES solutions.}
		\label{stats-fig6}
	\end{figure}
	
	We now compare turbulent quantities of the VES 1.35\% solution at MDR to XG 0.2\% and HPAM 0.05\% results in \hyperref[stats-fig6]{Figure~\ref*{stats-fig6}}, at the conditions indicated by the dashed square in \hyperref[stats-fig4]{Figure~\ref*{stats-fig4}}. As mentioned previously, the VES results are presented alongside well-known flexible (HPAM) and rigid (XG) polymer solutions at MDR. The streamwise Reynolds stresses are presented in \hyperref[stats-fig6]{Figure~\ref*{stats-fig6}} (a). Interestingly, the $\langle u^2 \rangle$ results for the VES are similar to the rigid polymer xanthan gum, where the general profile of $\langle u^2 \rangle$ in most of the duct as well as its peak value. In comparison to the VES and XG, the HPAM values are lower at $ 0.05 \leq y/h \leq 0.4$, and reach approximately the same values as the other drag-reducing fluids up to the centreline. 
	
	The recent study by \citet{warwaruk2021} compared a C14 surfactant and polyacrylamide (PAM) at MDR, where the $\langle u^2 \rangle$ profiles were nearly identical, at least for their dilute solutions. However, the distribution of streamwise Reynolds stress of HPAM and VES solution at MDR were quite different in our experiments. Since our experimental parameters for the MDR comparison show comparable Reynolds numbers and DR percentages in the data for all three viscoelastic fluids in \hyperref[stats-fig6]{figure~\ref*{stats-fig6}}, we can hypothesize that the difference in the $\langle u^2 \rangle$ profiles could be largely due to the shear viscosity characteristics of each fluid, since the effect of viscoelasticity at MDR should be mostly the same, as seen from the velocity profiles and friction factor data. For instance, the turbulent flow experiments by \citet{peixinho2005,mitishita2021} and simulations by \citet{singh2017b} with shear-thinning fluids noted larger $\langle u^2 \rangle$ peaks compared to Newtonian flows, an observation that the authors noted as a consequence of shear-thinning rheology. Therefore, spatial differences in viscosity (and thus in viscous stresses) along the duct could explain why the $\langle u^2 \rangle$ profiles of VES/XG are different from HPAM.
	
	There is little difference between each drag-reducing fluid in the profiles of $\langle v^2 \rangle$ in \hyperref[stats-fig6]{Figure~\ref*{stats-fig6}} (b) and $-\langle uv \rangle$ in  \hyperref[stats-fig6]{Figure~\ref*{stats-fig6}} (c). The results follow the trend at MDR where both wall-normal Reynolds stresses and Reynolds shear stresses are greatly reduced, correlating to the decrease of overall turbulence production in the flow due to the viscoelastic additives. The skewness profiles between the polymers and surfactants in \hyperref[stats-fig6]{figure~\ref*{stats-fig6}} (d) also show small differences overall, and the general distribution is common at MDR. Finally, we recall the discussion regarding the asyptotic limit of Zakin for the turbulent flow of surfactants. Recent experiments such as the PIV analysis by \citet{warwaruk2021}, the turbulent velocity profiles of surfactants at MDR \citep{warholic1999}, and the injection flow by \citep{tamano2010, tamano2018} show velocity profiles at MDR that closely follow the upper limit of Virk \citep{virk1970} of polymer solutions, instead of replicating Zakin's asymptote \citep{zakin1996} for surfactants at MDR. Our data turbulent flow data with wormlike micellar gels thus gives further evidence that the maximum drag reduction for the EDAB surfactants and perhaps other micellar fluids is likely limited by Virk's asymptote, and thus being equal to the upper limit of polymer DR, rather than Zakin's asymptote.
	
	\begin{figure}
		\centering
		\includegraphics*[width=140mm]{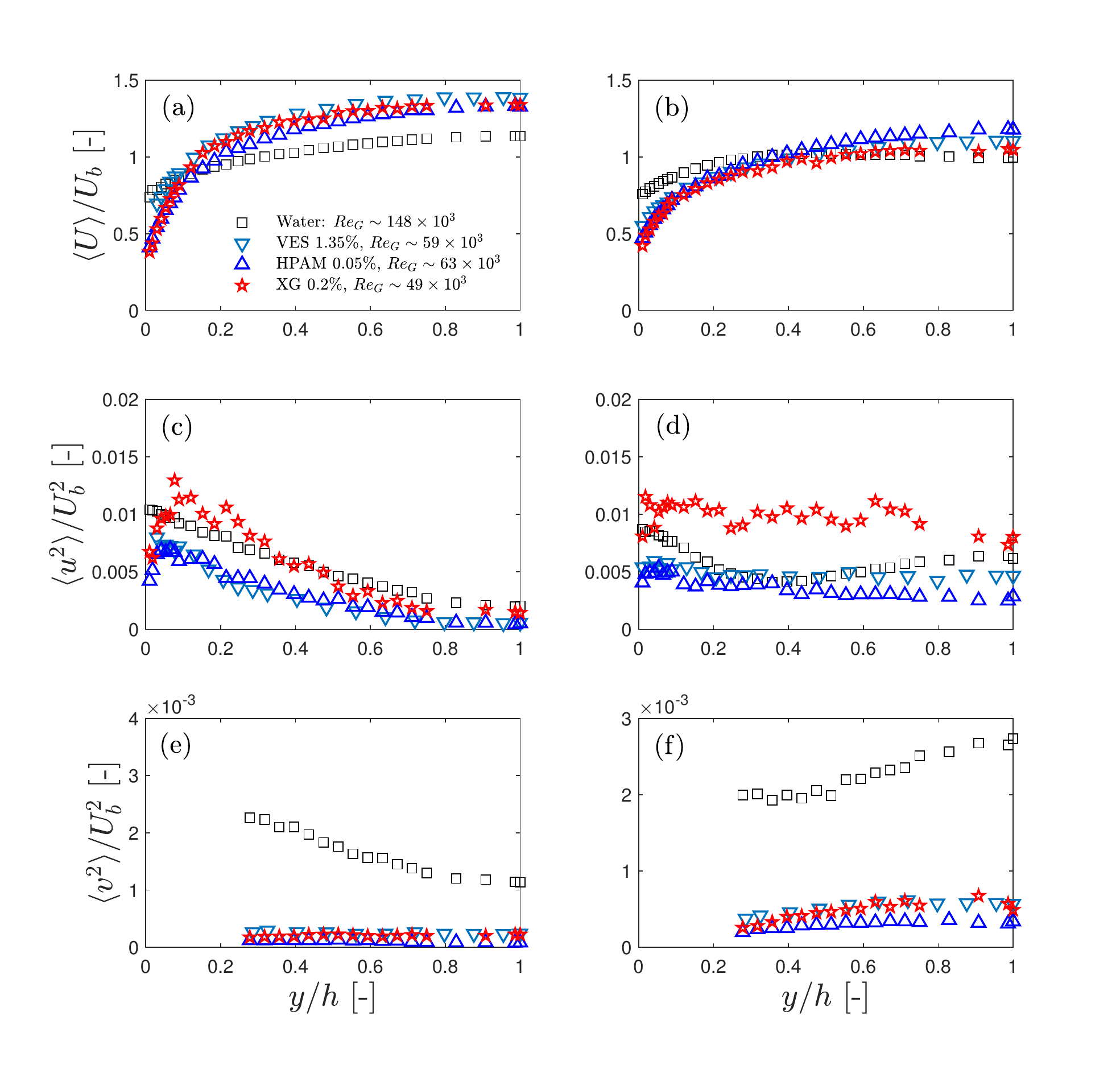}
		\caption{\fontsize{9}{9}\selectfont Velocity profiles (a-b), streamwise (c-d) and wall normal (e-f) Reynolds stresses of water, 0.05\% HPAM, 0.2\% XG and 1.35\% VES solutions. Measurements were performed at z/w = 0.2 and 0.6 (MP2 and MP3 positions in the rectangular duct).}
		\label{stats-fig7}
	\end{figure}
	
	For completeness, and to further characterize the turbulent flow dynamics in the duct, we measure the velocity, streamwise and wall-normal Reynolds stress profiles in two other different positions in the $z$-direction, MP2 ($z/w = 0.6$) and MP3 ($z/w = 0.2$) in \hyperref[flowloop2]{Figure~\ref*{flowloop2}}. The measurements are presented in \hyperref[stats-fig7]{Figure~\ref*{stats-fig7}}, wherein (a), (c) and (e) show $\langle U \rangle$, $\langle u^2 \rangle$, $\langle v^2 \rangle$ profiles at $z/w = 0.6$, respectively; letters (b), (d) and (f) show $\langle U \rangle$, $\langle u^2 \rangle$, $\langle v^2 \rangle$ profiles at $z/w = 0.3$, respectively. The velocity profiles at $z/w = 0.6$ are quite close to the duct centre-plane results from \hyperref[stats-fig1]{Figure~\ref*{stats-fig1}}, with the velocity profiles of polymer fluids and the VES being very close to each other. Moving closer to the side wall of the duct at $z/w = 0.2$, we observe a larger influence of secondary flows near the corners in the water flow, with a small a peak in the average velocity, and a flatter overall velocity profile of XG. The VES and HPAM velocity profiles are still approximately parabolic, and are quite close to each other.
	
	At $z/w = 0.6$ here we observe that the $\langle u^2 \rangle$ values of the VES are now approximately the same as the HPAM ones, in contrast to the trend seen in \hyperref[stats-fig6]{Figure~\ref*{stats-fig6}}(a) where the $\langle u^2 \rangle$ profile of the VES in the centre-plane was similar to XG. This indicates that, at MDR, the spanwise variation in $u$ fluctuations is different for each viscoelastic fluid investigated. At $z/w = 0.2$, the $\langle u^2 \rangle$ values of both HPAM and VES are significantly lower than water near the corners, while the XG shows much larger fluctuations across all $y/h$ positions. Our results show that the peak in $\langle u^2 \rangle$ of VES and HPAM fluids decreases as we move closer to the side wall, following the results of \citep{shahmardi2019} of DNS in a square duct with polymeric fluids. The $\langle v^2 \rangle$ show the same trend to what was observed in the centre-plane of the duct, with a large decrease $v$ fluctuations in the HPAM, XG and VES compared to water. There is no significant difference in $\langle v^2 \rangle$ when comparing all three viscoelastic solutions investigated. The $\langle v^2 \rangle$ increases slightly near the side wall at $z/w = 0.2$ when compared to the values at $z/w = 0.6$, again qualitatively in agreement to \citep{shahmardi2019}. These results show that even at MDR, when time-averaged velocity profiles in the centre-plane, $Re_G$ and friction factor values of VES, XG and HPAM are almost the same, the flow characteristics across the entirety of the duct show some differences, especially in the $\langle u^2 \rangle$ profiles.
	
	From the velocimetry results, the entangled gel-like structure of the VES in the near-wall region appears to be broken down in a fully turbulent flow, likely into smaller rods or even spherical micelles, at least in the concentrations investigated. For low DR cases such as with the VES 0.50\%, it is possible that the micellar structure is broken down everywhere in the duct because the lower concentration VES solutions cannot withstand high deformation rates (especially near the wall) as well as the higher concentration solutions. The breakdown of the wormlike micelles makes sense if we take into account that the dynamic breaking and reforming of micelles is a consequence of weak bonds that are easily undone by shear and extension in turbulence, unlike the strong covalent bonds from polymer molecules \citep{rothstein2008}. A recent polymer DR investigation by \citet{mohammadtabar2020} stated that linear rheology measurements cannot be correlated drag reduction, and it appears that the same is true for our micellar gels. From our results, TDR may be correlated to the micellar structure characteristic of the turbulent flow, which is likely completely different from the conditions in a rotational rheometer. This means that the properties of VES measured in the linear viscoelastic regime, such as the relaxation time, could be irrelevant for quantitative correlations to the turbulent regime.
	
	\subsection{Probability density functions} \label{s4.3}
	
\begin{figure}
		\centering
		\includegraphics*[width=140mm]{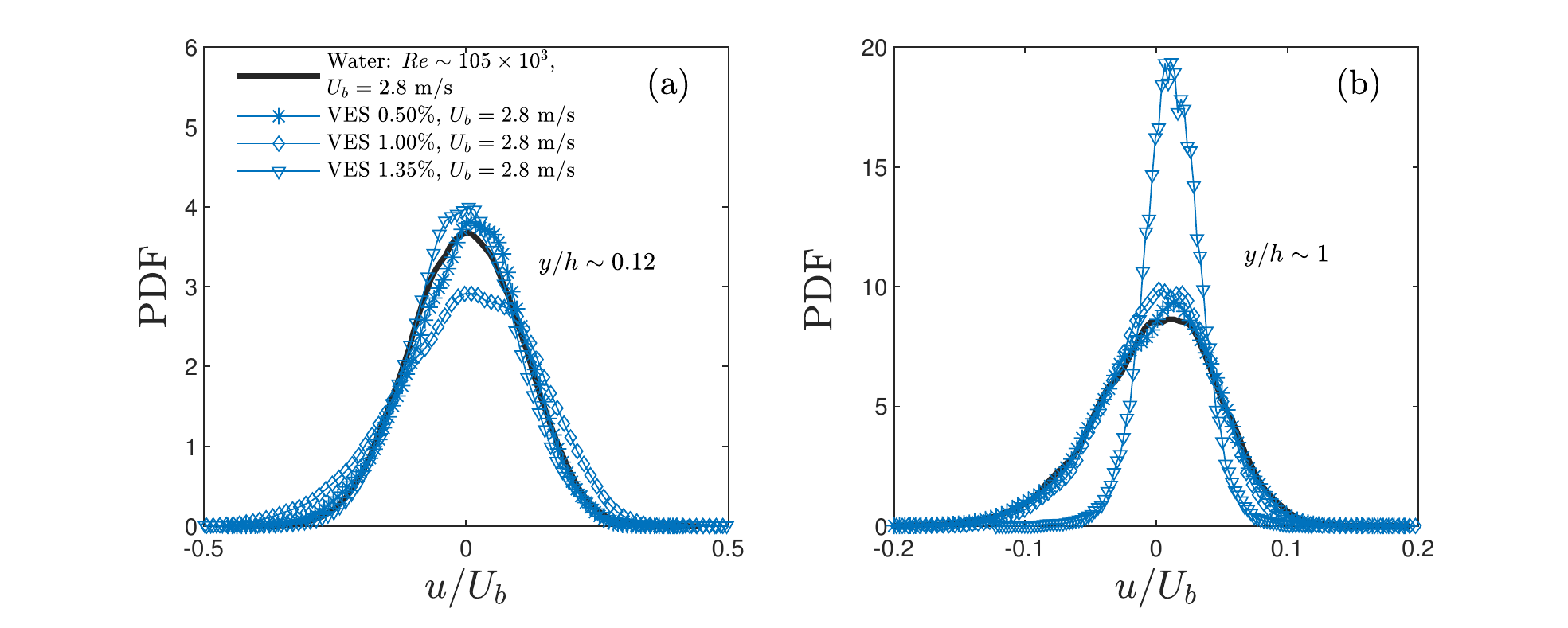}
		\caption{\fontsize{9}{9}\selectfont Probability density functions of water and 0.50\%, 1.00\% and 1.35\% VES solutions at $U_b = 2.8~m/s$.}
		\label{pdf-fig1}
	\end{figure}

	We investigate the turbulent structure of the VES flow via probability density functions (PDFs) of velocity fluctuations. We begin by presenting the PDFs of streamwise velocity fluctuations $u$ of all three concentrations of VES near the wall at $y/h = 0.12$ and centreline of the duct in \hyperref[pdf-fig1]{Figure~\ref*{pdf-fig1}} (a). Near the wall, the distribution of $u$ in water and the VES 0.50\% solutions is nearly the same. A wider distribution can be seen with VES at 1.00\% concentration, with larger skewness in the negative direction, as seen in \hyperref[stats-fig5]{Figure~\ref*{stats-fig5}} (d). The wider PDF of $u$ of VES 1.00\% correlates to the large peak in streamwise Reynolds stress observed in the same position. The MDR distribution of $u$ again resembles the water value due to the presence of the fact that the small peak in $\langle u^2 \rangle$ is positioned at $y/h = 0.12$. The situation in the centreline shown by \hyperref[pdf-fig1]{Figure~\ref*{pdf-fig1}} (b) is more interesting because of the large differences seen in the distributions of $u$ in the VES 1.00\% (HDR) and VES 1.35\% (MDR). Even though the \%DR are not all that different between the VES 1.00\% and 1.35\% solutions, (62\% \textit{vs.} 69\%), the probabilities of small fluctuations nearly doubles. The PDF of $u$ at MDR thus indicates very weak turbulence in the centreline of the duct.
	
\begin{figure}
		\centering
		\includegraphics*[width=120mm]{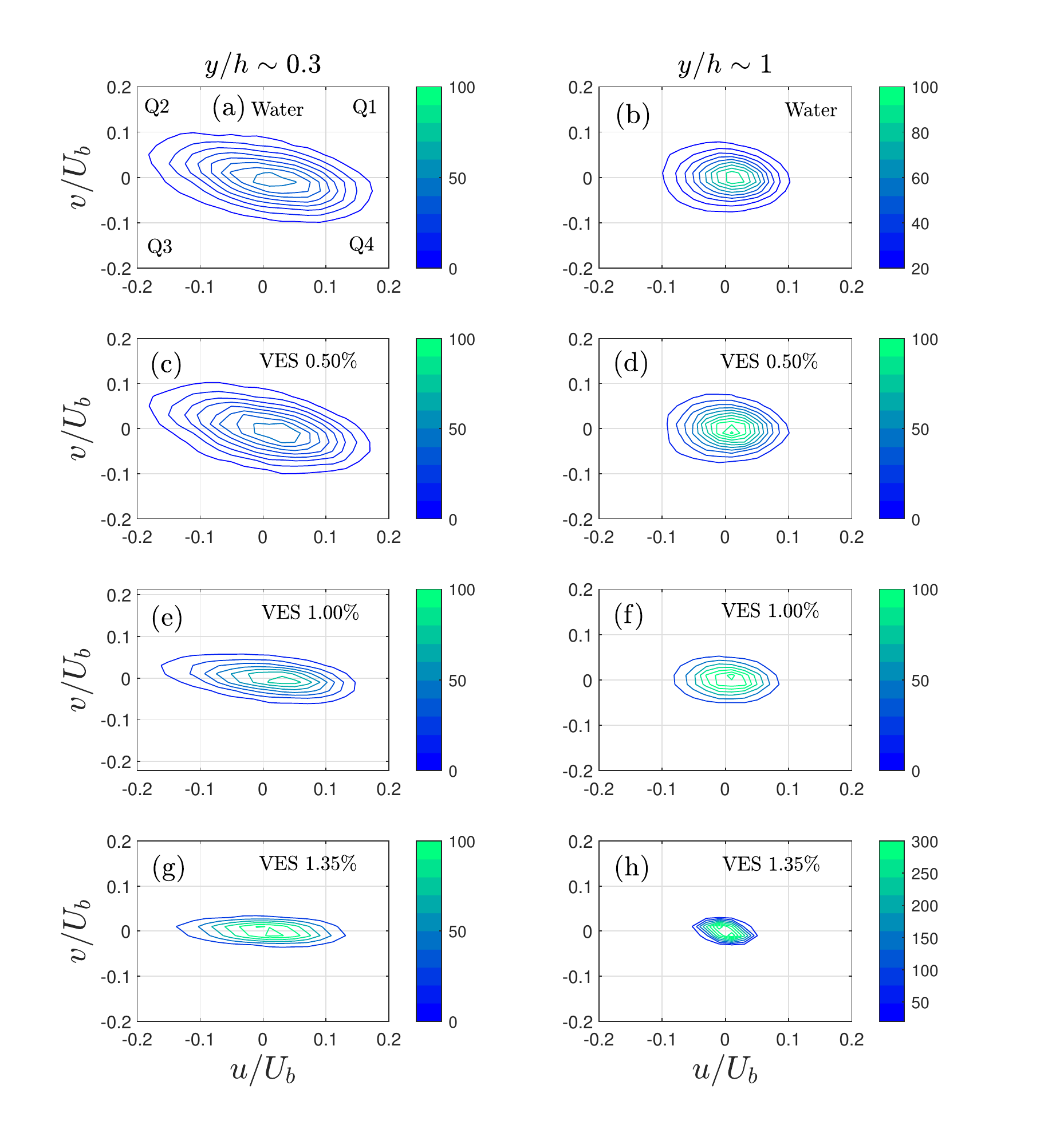}
		\caption{Joint probability density functions of water (a-b) and 0.50\% (c-d), 1.00\% (e-f) and 1.35\% (g-h) VES solutions measured at $y/h = 0.3$ and $y/h = 1$, at $U_b = 2.8~m/s$.}
		\label{pdf-fig2}
	\end{figure}

	We investigate the contributions of ejection and sweep motions to the turbulent flow via joint probability density functions (JPDF), analyzed in each quadrant (Q1 to Q4) of the $u$-$v$ space: $u > 0, v > 0$: Q1 events or outward interactions; $u < 0, v > 0$: Q2 events or ejections; $u < 0, v < 0$: Q3 events or wallward interactions and $u > 0, v < 0$: Q4 events or sweeps \citep{wallace1972, mohammadtabar2017}. \hyperref[pdf-fig2]{Figure~\ref*{pdf-fig2}} shows the JPDFs for water and the three VES concentrations at two positions in the duct: $y/h \sim 0.3$, which is the position nearest to the wall where we can perform coincident measurements of both $u$ and $v$, and the centreline $y/h \sim 1$. \hyperref[pdf-fig2]{Figure~\ref*{pdf-fig2}} (a) and (b) show results for water, (c) and (d), VES 0.50\%, (e) and (f), VES 1.00\% at HDR and (g) and (h), VES 1.35 at MDR\%, all at $U_b = 2.8~m/s$.
	
	We find that the JPDFs for the VES 0.50\% solution in \hyperref[pdf-fig2]{Figure~\ref*{pdf-fig2}} (c) and (d) are very close to water. We observe dominant ejection (Q2) and sweep events (Q4) in both water and VES 0.50\% results at $y/h \sim 0.3$, with small probabilities of low $u$ and $v$ fluctuations. The centreline results show less intense ejection and sweep motions, with a large probability density near zero values of $u$ and $v$. As we increase the VES concentration to 1.00\% (HDR regime) in \hyperref[pdf-fig2]{Figure~\ref*{pdf-fig2}} (e) and (f), the shapes of the JPDF become flatter at $y/h \sim 0.3$, with lower intensity of $v$ fluctuations. Compared to the VES 0.50\%, there is a small decrease in $u$ motions, but it is not as pronounced as the change in $v$. Ejection motions are still dominant over sweeps in HDR, but the probability of near-zero $u$ and $v$ fluctuations increases over the VES 0.50\%. At the centreline, $u$ decreases drastically, while $v$ fluctuations remain approximately the same as the $y/h \sim 0.3$ position. With the VES at 1.00\% concentration, the overall reduction $-\langle uv \rangle$ near the wall contributes to a largely less turbulent core, dominated by mostly weakened streamwise structures. As seen in \hyperref[stats-fig5]{figure~\ref*{stats-fig5}}, $\langle u^2 \rangle > \langle v^2 \rangle$ and the $\langle v^2 \rangle$ profile is much lower than in a Newtonian flow, which was also verified in DNS simulations of turbulent drag reduction \citep{xi2010,pereira2017b}. 
	
	At MDR the pattern of the JPDFs of $u$ and $v$ fluctuations in \hyperref[pdf-fig2]{Figure~\ref*{pdf-fig2}} (g) and (h) is markedly different from the VES 0.50\% and water at $y/h \sim 0.3$. The $v$ fluctuations are greatly reduced in comparison to the water flow, and the near-wall turbulence is mostly dominated by streamwise fluctuations, with a JPDF that is ``flatter" than the VES 1.00\% result. The fluctuations in the centreline decrease further when compared to VES 1.00\%, where we see a very large probability density of near-zero $u$ and $v$ fluctuations. The energy balance analysis by \citep{hara2017} with an MDR flow of a dilute CTAC solution appears to be valid here. They stated that the elastic energy from the wormlike micelles, released in the buffer layer, is the main reason why turbulence is sustained at MDR, especially when turbulence production is negligible as seen from the low $-\langle uv \rangle$. Therefore, this net energy transfer results in overall low values of $u$ fluctuations. The DNS study by \citet{pereira2017b} with the FENE-P model proposed a similar mechanism for polymer solutions, where polymer molecules extract energy from the regions farther from the wall, and release energy to vortex structures in the near-wall region.  
	
\begin{figure}
		\centering
		\includegraphics*[width=120mm]{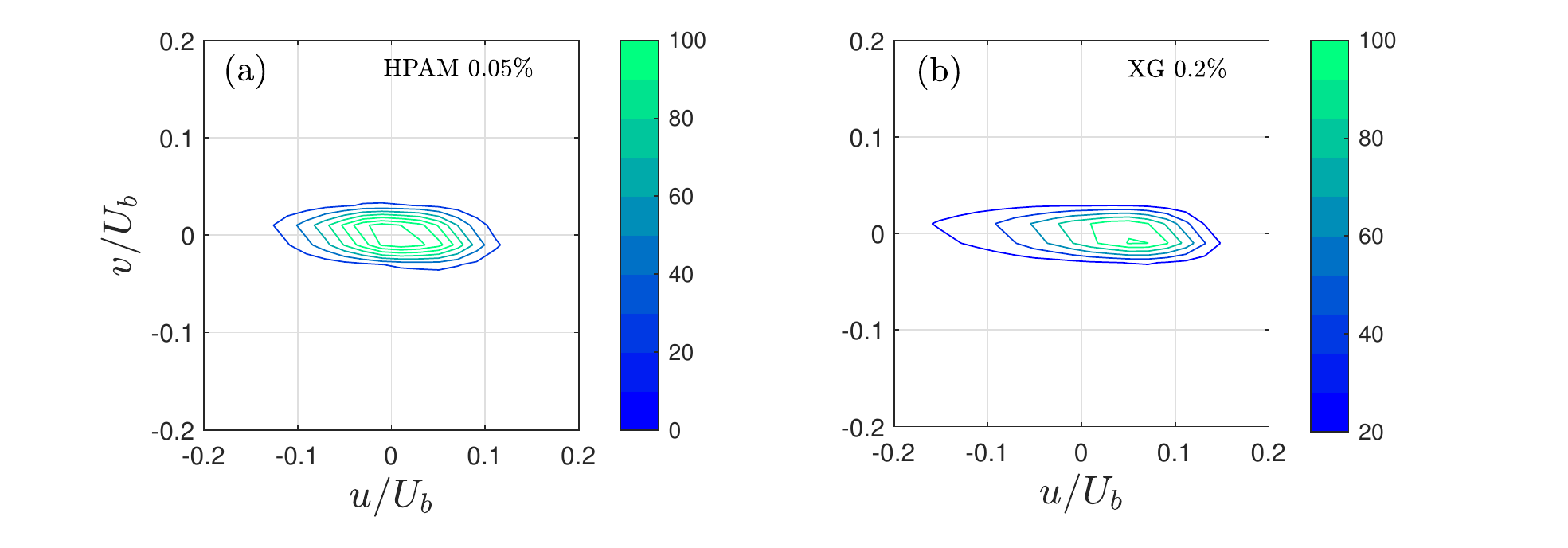}
		\caption{\fontsize{9}{9}\selectfont Joint probability density functions of 0.05\% HPAM, 0.2\% XG solutions measured at $y/h = 0.3$, $U_b = 3.8~m/s$.}
		\label{pdf-fig3}
	\end{figure}

	The results from \hyperref[pdf-fig2]{Figure~\ref*{pdf-fig2}}, analyzed in conjunction with the JPDFs of the polymer solutions at MDR, in the position $y/h \sim 0.3$ in \hyperref[pdf-fig3]{Figure~\ref*{pdf-fig3}}, further corroborate the conclusion that the mechanisms drag reduction of polymers and our VES appear to be alike, at least in the MDR regime. Both \hyperref[pdf-fig2]{Figure~\ref*{pdf-fig2}} (g) and \hyperref[pdf-fig3]{Figure~\ref*{pdf-fig3}} (HPAM and XG solutions) present JPDFs of $u$ and $v$ fluctuations that are qualitatively very close to each other, in the sense that the flow is mostly dominated by enhanced streamwise fluctuations at $y/h \sim 0.3$. The JPDFs near the centreline are also nearly the same for both polymers and VES at MDR (not shown), with comparable turbulence quantities. We note that our study is limited to comparison to polymer flows in the MDR regime only, and data from the LDR (\%DR $< 40$) and HDR (\%DR $> 40$) regimes with lower concentrations could be useful to broaden the scope of this study. However, the effect of VES concentration in the JPDFs at different \%DR is at least qualitatively similar to xanthan gum solutions in a turbulent channel flow \citep{mohammadtabar2017}.
	
	\subsection{Streamwise power spectral densities} \label{s4.4}
	
	We investigate the one-dimensional power spectral densities (PSDs), represented by $E_{uu}$ for streamwise velocity fluctuations $u(t)$. We define the power spectral density as an estimate of the energy distribution throughout the frequency range, i.e.~$E_{uu} \propto |u_f(f)|^2$, where $u_f(f)$ is the Fourier transform of $u(t)$, for a given $y/h$ position. We convert the frequency data $f$ into the wavenumber space by $k_x = 2 \pi f / \langle U \rangle$. The wavenumber is then made dimensionless by multiplying by the duct half-height $h$. This is possible by using Taylor's hypothesis \citep{pope2001}, which is valid if the root mean square of the velocity fluctuations $u_{rms}$ is less than 20\% of the mean local velocity $\langle U \rangle$ \citep{tran2010}. The PSD is made dimensionless dividing by $\nu_s U_b$, where $\nu_s$ is the kinematic viscosity of the solvent. 
	
\begin{figure}
		\centering
		\includegraphics*[width=140mm]{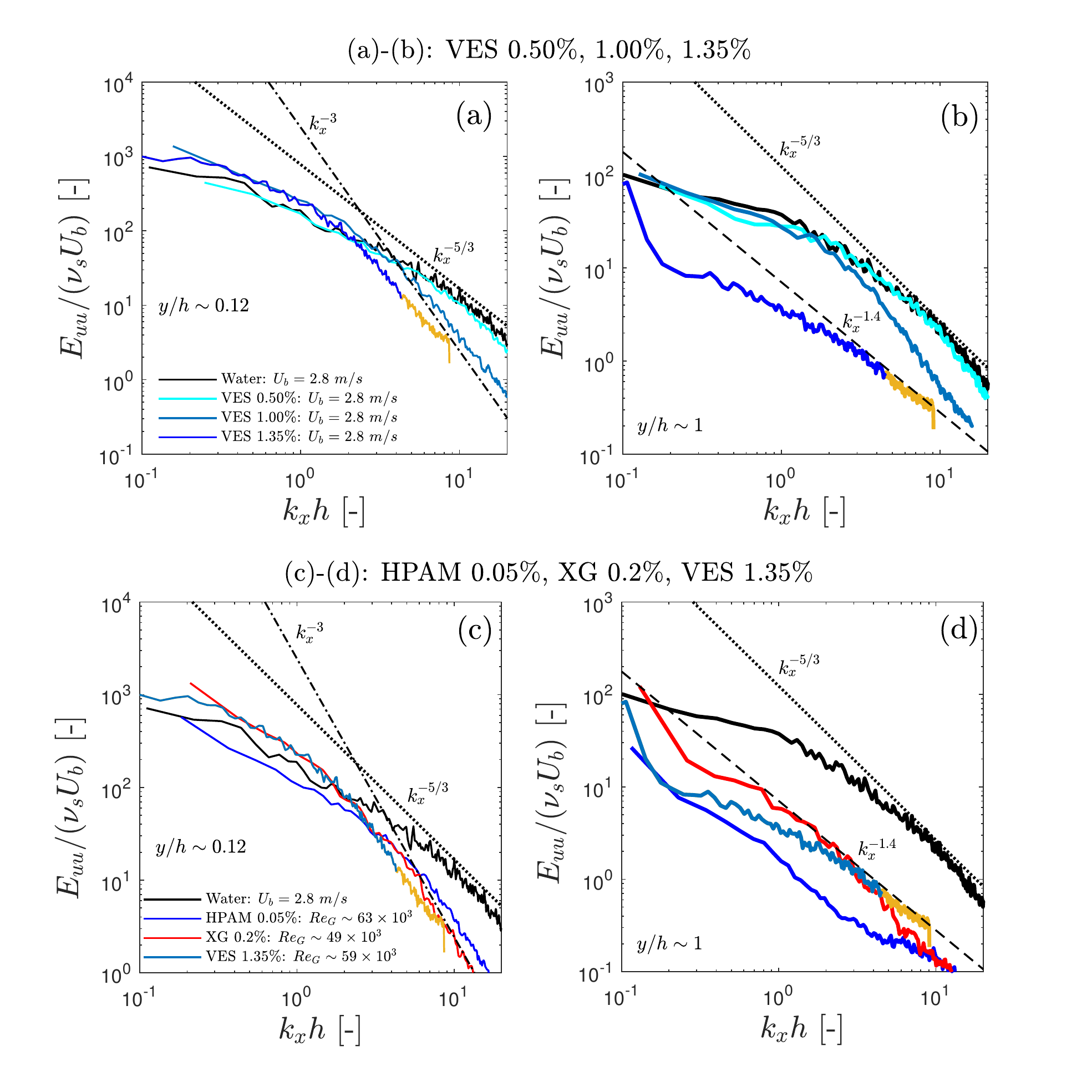}
		\caption{\fontsize{9}{9}\selectfont Streamwise power spectral densities of 0.50, 1.00 and 1.35\% VES solutions measured at $y/h = 0.12$ (a) and $y/h = 1$ (b). The power spectral densities of 0.05\% HPAM, 0.2\% XG compared to the 1.35\% VES at $y/h = 0.12$ and $y/h = 1$ in the MDR state are shown in (c) and (d) respectively. Water PSDs are presented for comparison, along with the $k_x^{-5/3}$ slope for the inertial range in dotted lines. Yellow lines represent PSD data without the cut-off limit. The slopes $\sim k_x^{-3}$ in dot-dashed lines and $\sim k_x^{-1.4}$ in dashed lines are merely guides to the eye to represent the energy decay of the drag-reduced flows with wavenumber.}
		\label{spec-fig1}
	\end{figure}

	It is not possible to obtain equally spaced velocity measurements in time with LDA because each data point depends on the particles to pass through the measurement volume. Thus, we perform a linear interpolation of the velocity-time signal to obtain equally spaced data points. The interpolation frequency is the average data rate of the experiments \citep{toonder1997} which varies between 1000 and 4000 \textit{Hz} for the HPAM, XG, VES 0.50\% and VES 1.00\% fluids. The VES 1.35\% solution was slightly turbid even after heating, and we could only obtain an average data rate between 500 and 1500 \textit{Hz}. The interpolation acts as a filter to the data signal at high frequencies, and therefore we cut off our results at a maximum frequency of approximately 1/4 of the total data rate, which is an adequate approximation according to \citet{ramond2000}, and does not exclude much of the high wavenumber range. However, the VES 1.35\% data was severely limited by this approach because of the lower data rates, so we present both the PSDs with the wavenumber limiter, and also without the cut-off limit as an approximation for the energy at high wavenumber, seen in yellow. The accuracy of our PSD measurements with water has been benchmarked in \citet{mitishita2021}, with good agreement with Kolmogorov's -5/3 scaling for $E_{uu}$ in inertial range. The PSDs of water from  \citep{mitishita2021} are presented alongside the VES and polymer results for reference. The PSDs for all concentrations of VES fluids are presented in \hyperref[spec-fig1]{Figure~\ref*{spec-fig1}}, where (a) and (b) show the PSDs for the VES 0.50\%, 1.00\% and 1.35\% at $y/h \sim 0.12$ and $y/h \sim 1$, respectively. \hyperref[spec-fig1]{Figure~\ref*{spec-fig1}} (c) and (d) compare the PSDs of the polymers HPAM 0.05\% and XG 0.2\%, and VES 1.35\% (recall the dashed rectangle indicating the fluid parameters in \hyperref[stats-fig4]{Figure~\ref*{stats-fig4}}) at the same positions as (a) and (b).
	
	We observe that the PSDs of VES 0.50\% and water are nearly identical, at least in the positions investigated in \hyperref[spec-fig1]{Figure~\ref*{spec-fig1}} (a) and (b), following the -5/3 slope for $E_{uu}$ in the inertial range. At low \%DR, \citet{warholic1999influence} observed that the $E_{uu}$ in turbulent flows of water and a polyacrylamide polymer solution at low concentration were nearly the same. The spectral results thus correlate to our previous claim that the wormlike micelles in the VES 0.50\% are mostly broken down into smaller rods or possibly spherical micelles. 
	The VES 1.00\% results qualitatively agree to the data from \citet{warholic1999}, with a drop in spectral energy seen at high wavenumbers as $k_x h > 4$ with a slope of $\sim k_x^{-3}$, in our case. Interestingly, there is a small region of $E_{uu}$, for $k_x h < 4$ where we observe a small section with the $k_x^{-5/3}$ slope. In our previous work \citep{mitishita2021}, the energy spectra of turbulent flow of Carbopol solutions with mostly shear-thinning effects at $y/h \sim 0.12$ showed enhanced energy at low wavenumbers, whereas in a drag-reducing flow the energy content at low wavenumbers was nearly the same as water. The VES 1.35\% data is quite limited here due to the low data rates from the LDA, but it does appear that the $k_x^{-5/3}$ slope in the inertial range has disappeared completely, and now the energy decreases with $\sim k_x^{-3}$ as well. In the case of MDR near the wall, the energy content drops earlier than VES 1.00\%, also with slope of $\sim k_x^{-3}$, near $k_x h \sim 1$.
	
	Now at the centreline, the PSDs of VES 0.50\% and VES 1.00\% follow the same trend as \hyperref[spec-fig1]{Figure~\ref*{spec-fig1}} (a), with the $E_{uu}$ curve of water being identical as the VES 0.50\% and very close to a $k_x^{-5/3}$ slope, and the $E_{uu}$ curve of VES 1.00\% being nearly the same as water until $k_x h \sim 4$, where we see a drop in energy content of higher wavenumbers with $k_x^{-3}$. However, the PSD of VES 1.35\% is very different in the centreline. We observe a large decrease in the total energy content of all wavenumbers, with $E_{uu}$ decreasing with a slope of $\sim k_x^{-1.4}$. Even when considering our experimental limitations of acquisition rate, the data is consistent with the previous JPDF data of \hyperref[pdf-fig2]{Figure~\ref*{pdf-fig2}} (g), where it showed a large probability density of very small (near-zero) $u$ and $v$ fluctuations. This means that the core is significantly less turbulent at MDR, with the majority of the turbulent kinetic energy near the walls. 
	
	The picture of MDR for polymers and surfactants is very similar in terms of energy spectra, as seen in \hyperref[spec-fig1]{Figure~\ref*{spec-fig1}} (c) and (d). Near the wall, the $E_{uu}$ curves from polymers and VES are close to water at low $k_x$, and then decrease below water near $k_x h \sim 4$ with a slope that follows $k_x^{-3}$. A recent work on turbulent polymer jets by \citet{yamani2021} also observed a $f^{-3}$ slope (note that frequency is analogous to wavebumber) in the power spectra of fluctuations of local polymer concentrations. The spectra results were shown to be independent of polymer molecular weight and concentration. With dimensional arguments from \citep{hinch1977}, they associated the $f^{-3}$ slope to the time-averaged strain-rate that is dominant in elastoinertial turbulence, which is dominant in HDR or MDR flows \citep{choueiri2018,xi2019}. 
	
	Regarding our data, we note that more quantitative analyses such as scaling laws and energy dissipation estimates are difficult due to experimental data rate limitations, where $E_{uu}$ curves spanning many decades in $k_x$ cannot be computed. The low data rates are quite evident in the limited VES 1.35\% spectra, for instance. Therefore, the scaling relationships with $k_x$ for drag-reduced flows are merely guides to the eye in our results due to these experimental limitations, as they are not well-established theoretically like Kolmogorov's $k_x^{-5/3}$ scaling. Thus, more duct flow experiments need to be performed with polymers at different concentrations at an increased acquisition rate with the LDA, in an attempt to better analyze the $\sim k_x^{-3}$ slope of $E_{uu}$ with our setup. The observation of the $k_x^{-3}$ slope is nevertheless interesting and warrants further study. The centreline PSDs for the polymers are somewhat in agreement to the VES solutions, with a significant decrease in energy in all $k_x$ values.
	
	To finish the discussion of our experimental results, we summarize the effects of $Re_G$ in the turbulent flow of a VES 1.35\% solution, which is able to reach MDR at $U_b = 2.8~m/s$. An extended presentation of this data is given in the Supplementary Material for completeness. The velocity data suggest that a high-$Re_G$ flow is able to reduce the micellar structure of VES from entangled worms, to shorter rods, and spherical micelles. It is analogous to a progressive decrease in concentration due to the dynamic nature of the micellar structure \citep{cates1990c,dreiss2017}. In contrast to Carbopol solutions for example, the EDAB/VES fluids are not permanent gels, but viscoelastic fluids with very long relaxation times in the linear viecoelastic regime \citep{raghavan2001, chu2010, gupta2021}, and in this context it makes sense to assume that a fully turbulent flow would completely break down a micellar gel. The breaking of wormlike micelles near the wall with larger $Re_G$ reduces the amount of energy absorbed from the turbulent structures near in the core or log layer \citep{hara2017,pereira2017b,pereira2018}. The streamwise PSDs are further evidence of the approach to water flow as $Re_G$ increases, where the energy spectra approaches the $k_x^{-5/3}$ line for the inertial range of Newtonian fluids. Recently, capillary flow rheometry has shown promise in quantifying the relationship between micellar structure and surfactant rheology, such as in the experiments by \citep{salipante2020} where they correlate micellar length scales with the shear-thinning behaviour of surfactant solutions at high shear rates. Such an investigation with our VES/EDAB solution could then be useful in conjunction with turbulent flow data at large Reynolds numbers. Therefore, a high $Re_G$ flow decreases the size/length of the micelles in almost the same way as reducing its concentration, and the drag reduction mechanism becomes less effective.

	\section{Conclusion} \label{s6}
	
	We performed an experimental investigation of turbulent flows with a long-chained, zwitterionic surfactant solution (VES) at three different concentrations (0.05\%, 1.00\% and 1.35\%) and $Re_G$ values. All solutions were formulated in the semi-dilute regime, which endowed a shear-thinning rheology to all fluids. LDA experiments were performed to assess the effect of the VES concentration in the velocity and Reynolds stress measurements. Our data showed that only the VES 1.35\% solution was able to reach maximum drag reduction at $U_b = 2.8~m/s$. The VES 1.35\%  solution at MDR is compared to the turbulent flow of semi-dilute polymer solutions: a XG solution at 0.2\%w concentration, and an HPAM solution at 0.05\% concentration. The comparison between polymers and VES was made at MDR with a similar value of $Re_G$. 
	
	Flow curves from measurements of ramp-down experiments of pre-sheared VES without a rest period resulted in a better estimate of the near-wall viscosity for turbulent duct flow conditions. Specifically for the VES at 1.35\% concentration, the $U^+$ \textit{vs} $y^+$ matched the velocity measurements of polymers at MDR. The results implied that the viscosity reduction in the VES due to shear is larger in the duct than what was measured by the rheometer. We believe this is a consequence of larger breakdown of the micelles in the near-wall regions of the duct when compared to the laminar flow conditions in the rheometer, which results in incorrect $y^+$ values.
	
	The JPDF and PSD results for the VES 0.50\% solution were almost identical to water in all duct positions shown here, with the energy spectra in particular showing a $k_x^{-5/3}$ scaling, identical to the Newtonian inertial range. It appears that a VES solution broken down by turbulence loses its gel-like properties seen in the rheometry experiments, and shows turbulent quantities of the same magnitude as the solvent. However, the fact that the \%DR was not zero means that there is still some energy exchange between the micelles and turbulent structures. Larger VES concentrations resulted in more drag reduction, a decrease in friction factor and Reynolds stresses. For a VES at 1.35\% concentration, the wormlike micelles appear to be broken down very near the wall, but less so farther away from it where the micelles can interact with turbulence.
	
	The comparisons of drag reduction with semi-dilute polymers and the gel-like VES solution show very similar turbulence statistics between the semi-dilute polymers and VES at MDR. A few differences between polymers and the VES at MDR are seen in the streamwise Reynolds stresses ($\langle u^2 \rangle$), where viscous stresses could be playing a larger role due to perhaps alignment of surfactant or polymer molecules. Nevertheless, the JPDFs show that the turbulent dynamics of polymer and wormlike micellar fluids at MDR have more simlarities than differences, with enhanced $u$ fluctuations and decreased $v$ fluctuations near the wall in all fluids tested. Power spectral densities of streamwise velocity fluctuations at MDR show only small-scale turbulent structures (large $k_x$) are weakened by polymers or VES in the near-wall region. In the centreline, turbulent structures of all length scales are substantially weakened. 
	
	The comparisons between polymer and VES solutions suggest a similar DR mechanism between the VES/EDAB and the HPAM and XG solutions, considering that our fluids did not appear to form SIS according to the rheology experiments. The gel-like rheological behaviour of the VES in the linear regime does not appear to affect the turbulent flow behaviour, because as mentioned by previous studies such as \citet{kumar2007} and more recently in \citet{gupta2021}, the VES is a non-permanent gel, and its relatively weak intermolecular bonds break down easily in turbulence. Therefore, the VES solutions generally behave akin to polymer solutions as concentration increases, in comparison to other experimental channel flow results such as \citep{escudier2009a, shaban2018}. Futher evidence of this is the fact that our results are at least in qualitative agreement with recent experiments on surfactant drag reduction \citep{tamano2018, warwaruk2021} at MDR, where surfactant drag reduction appears to be limited at the same MDR limit of polymer solutions, which is Virk's asymptote, rather than Zakin's. 
	
	To improve the present VES-polymer comparisons, additional data sets on polymer drag reduction are required, such low-DR and high-DR turbulent quantities. Detailed extensional rheology experiments could also be beneficial, but would likely suffer from the same limitations from the rheological tests in this paper. Flow conditions and micellar structure are likely going to be very different in a high-$Re_G$ duct flow when compared to experimental conditions in CaBER due to near-wall destruction of micellar structure, for instance. Therefore, measurements of extensional relaxation times of VES with CaBER could be misleading for TDR due to distinct micelle structure in a turbulent duct flow experiment. Nevertheless, the extensional viscosity of EDAB/VES has not yet been studied in detail, and could pave the road for additional discussion for turbulent drag reduction with wormlike micellar gels. 
	
	\section*{Acknowledgments} \label{Ackn}
	
	This research was made possible by funding from Schlumberger and NSERC under the CRD program, project 505549-16. Experimental infrastructure was funded by the Canada Foundation for Innovation and the BC Knowledge Fund, grant number CFI JELF 36069. This funding is gratefully acknowledged. We kindly thank Schlumberger for providing us the J590 surfactant, MI Swaco for supplying the HPAM polymer and CP Kelco for the donation of the xanthan gum used in this paper. R.S.M also acknowledges financial support from the University of British Columbia 4-Year Fellowship PhD scholarship program.
	
	\section*{Conflict of interests} \label{int}
	
	The authors report no conflict of interest.

	
	\bibliography{EDAB_drag_reduction}
	
\end{document}


\medskip  
		
\begin{center}
	{ \LARGE Supplementary material to the paper "Turbulent drag reduction of viscoelastic wormlike micellar gels"}
\end{center}


\section{Effect of the Reynolds number in polymer solutions at MDR}

\begin{figure}
\centerline{\includegraphics*[width=140mm]{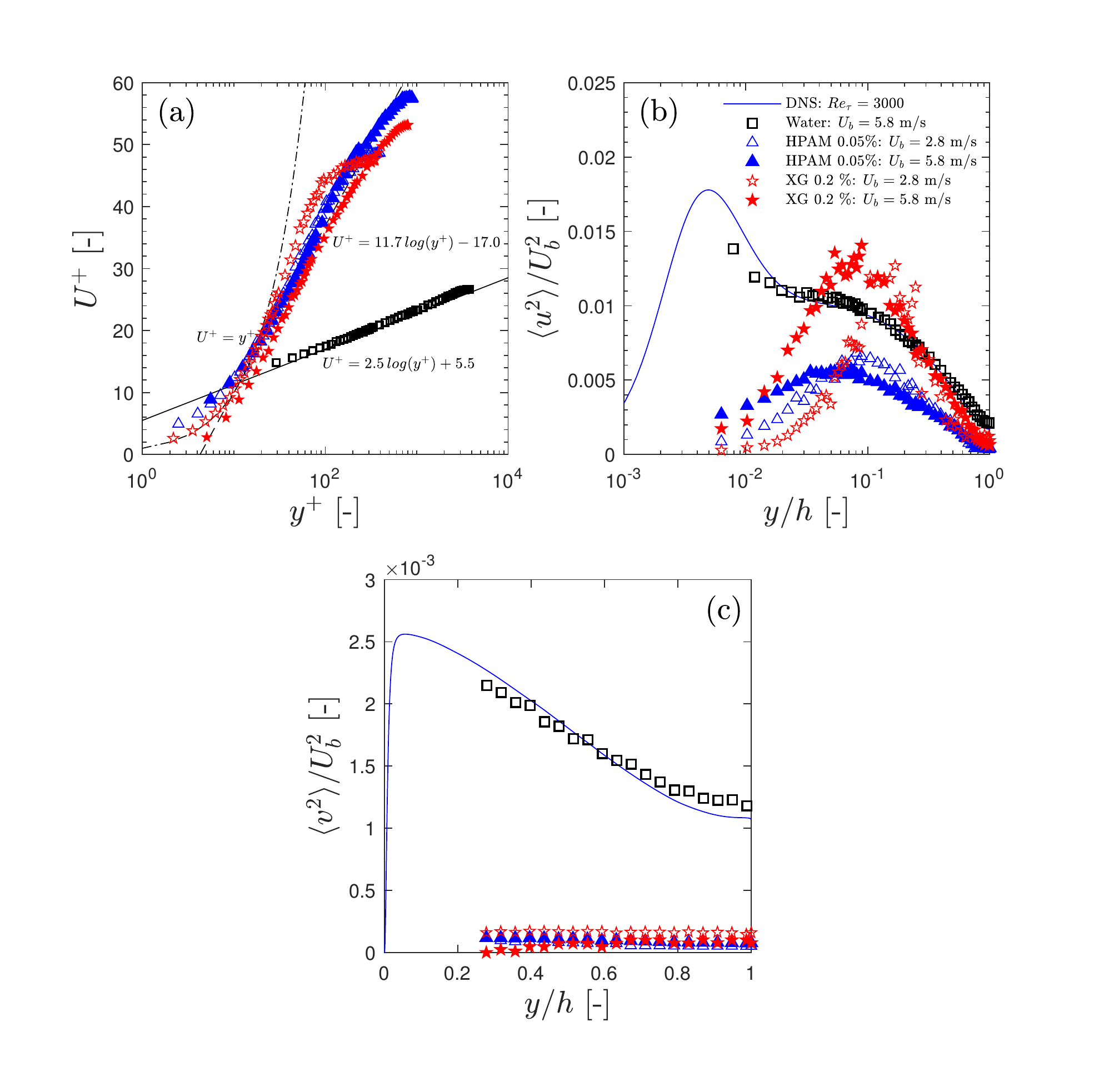} }
\caption{Inner scaled velocity profiles (a), and streamwise (b) and wall normal Reynolds stresses (c) HPAM and XG solutions at $U_b = 2.8$ and $5.8~m/s$.}
\label{sup-fig1}
\end{figure}

The turbulent drag reduction data at maximum drag reduction (MDR) with HPAM 0.05\% and XG 0.2\% at distinct $Re_G$ values is presented in \hyperref[sup-fig1]{figure~\ref*{sup-fig1}}. As expected, the velocity profiles presented in \hyperref[sup-fig1]{figure~\ref*{sup-fig1}} (a) with both XG 0.2\% and HPAM 0.05\% solutions are very similar, with data that is very close to Virk's asymptote for polymer drag reduction. Slight differences are shown in the XG profiles, however, where XG appears to exceed Virk's asymptote. We remind the reader that the MDR velocity line is purely empirical and therefore an average of many recorded experiments. The effect of $Re_G$ in the $\langle u^2 \rangle$ profile of HPAM 0.05\% in \hyperref[sup-fig1]{figure~\ref*{sup-fig1}} (b) is also quite minimal, and overall the profiles look very similar regardless of $Re$. Conversely, as $Re_G$ increases, the $\langle u^2 \rangle$ peak also increases and moves closer to the wall. We may hypothesize that, along with the changes in $U^+$ in the XG 0.2\% profile, is probably due to viscous effects because of the much higher shear rates in the duct flow as we increase $U_b$ from 2.8 to 5.8 \textit{m/s}. A similar effect in $\langle u^2 \rangle$ was observed in an inelastic, shear-thinning Carbopol solution in a turbulent flow, as reported by \citet{mitishita2021}. Finally, the effect of $Re_G$ in the $\langle v^2 \rangle$ profiles at MDR was as expected, with very low values of $\langle v^2 \rangle$ throughout the duct, which is characteristic of a flow at MDR.

\section{Effect of the Reynolds number in VES solutions}

\begin{figure}
	\centering
	\includegraphics*[width=100mm]{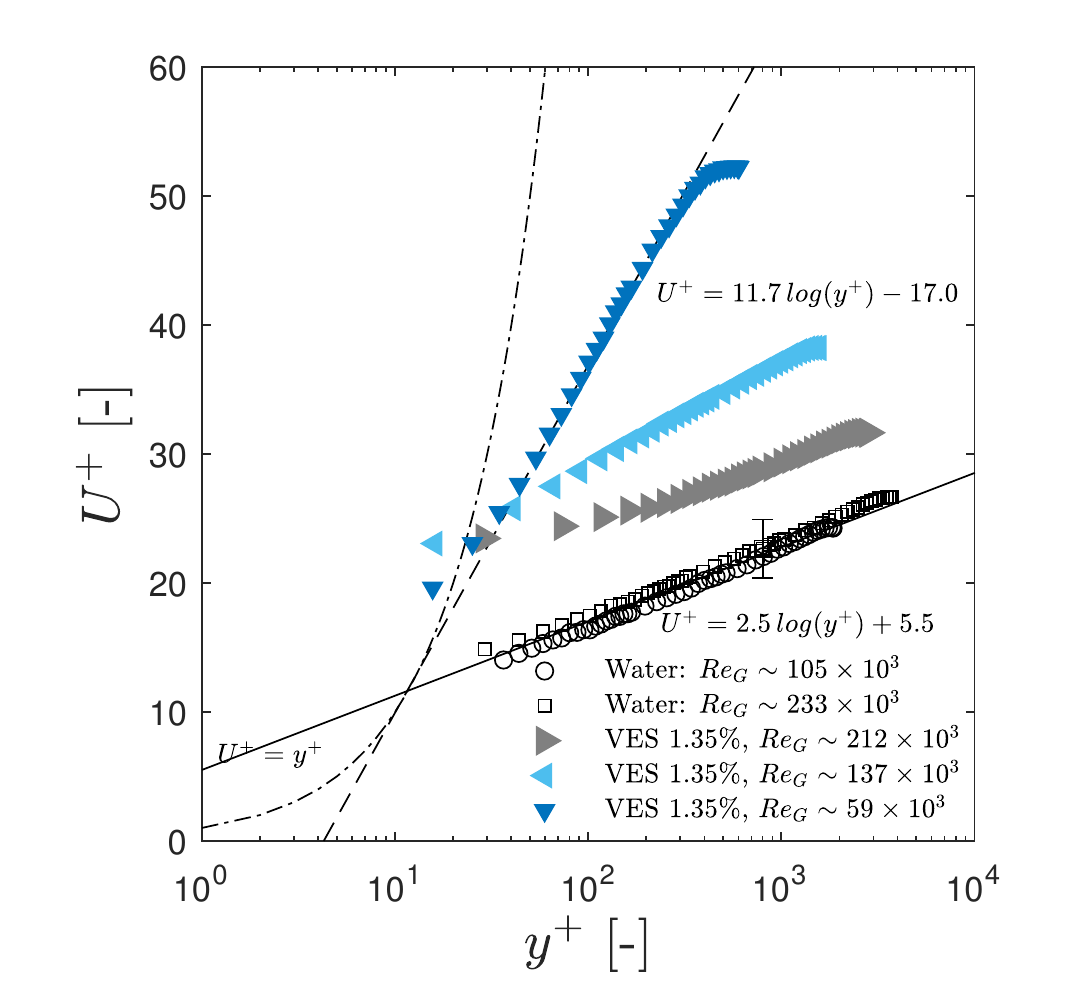}
	\caption{\fontsize{9}{9}\selectfont Velocity profiles of water and 1.35\% VES at $U_b = 2.8,~3.8$ and $5.8~m/s$ ($Re_G \sim 59 \times 10^3$, $137 \times 10^3$ and $212\times 10^3$ respectively), normalized in wall units.}
	\label{degr-fig1}
\end{figure}

	
	In this section we provide more details on the effects of $Re_G$ in the turbulent quantities measured by LDA. To this end, we investigate the turbulent flow of VES 1.35\% at $U_b = 2.8,~3.8$ and $5.8~m/s$, corresponding to $Re_G$ values of $\sim 59 \times 10^3$, $137 \times 10^3$ and $212\times 10^3$, respectively. The velocity and Reynolds number data, along with the parameters used to define $U^+$ and $y^+$ are listed in the main paper. We begin the analysis with the inner-scaled velocity profiles $U^+$ against $y^+$ in \hyperref[degr-fig1]{Figure~\ref*{degr-fig1}}. 

The data of the VES 1.35\% at MDR is the same as the one presented in the inner-scaled velocity profiles of the main paper, shown alongside the velocity profiles for the same solution at higher velocities. The increase in $U_b$ (or $Re_G$) causes a significant decrease in drag reduction, as seen in the velocity profiles of VES 1.35\% at $Re_G \sim 137 \times 10^3$, where it is in the high drag reduction regime (HDR, with DR $>$ 40\%). Increasing the velocity to $Re_G \sim 212 \times 10^3$ further lowers the drag reduction, as evidenced by the lower $U^+$ values. The velocity profiles in both VES flows apart from MDR are approximately parallel to the log law but shifted upwards. Note that the trend in the velocity profile of VES 1.35\% at $Re_G \sim 212 \times 10^3$ appears change near the wall, likely due to the fact that it becomes increasingly difficult to accurately measure near wall velocity with LDA as the flow velocity increases.

\begin{figure}
	\centering
	\includegraphics*[width=140mm]{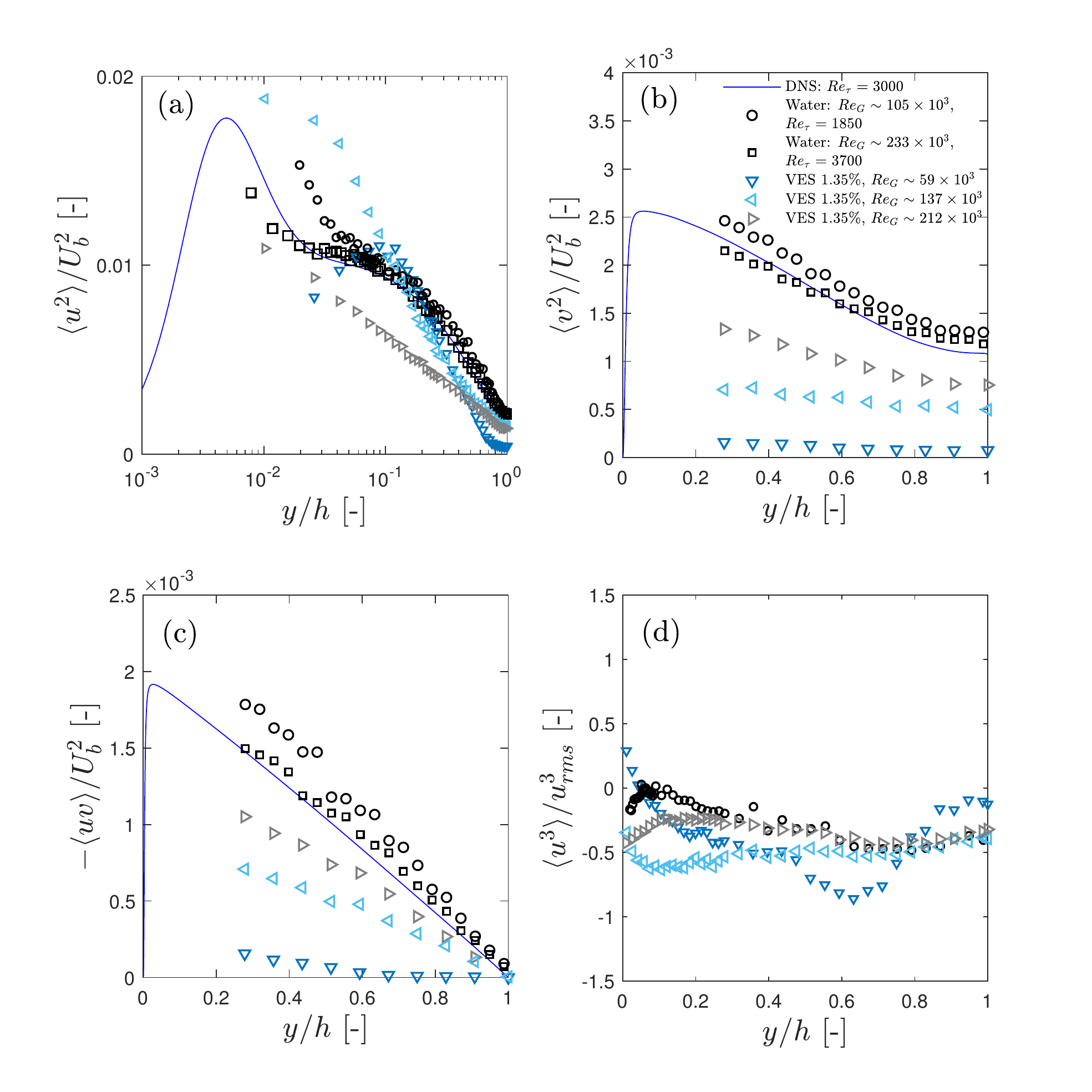}
	\caption{\fontsize{9}{9}\selectfont Streamwise (a), wall normal (b), shear Reynolds stresses (c) and skewness (d) of streamwise fluctuations of 1.35\% VES at $\sim 59 \times 10^3$, $137 \times 10^3$ and $212\times 10^3$.}
	\label{degr-fig2}
\end{figure}

The Reynolds stresses are presented in \hyperref[degr-fig2]{Figure~\ref*{degr-fig2}}. As we increase $Re_G$ from $59 \times 10^3$ to $137 \times 10^3$ the $\langle u^2 \rangle$ profile in \hyperref[degr-fig2]{Figure~\ref*{degr-fig2}} (a) becomes quite similar to the Reynolds stress profile of VES 1.00\% in the main paper. The peak in $\langle u^2 \rangle$ moves closer to the wall and increases substantially compared to the MDR case, to the order of the Newtonian DNS. The peak is expected to move even closer to the wall as we increase $Re_G$ to $212 \times 10^3$. However, if we compare the $\langle u^2 \rangle$ results from VES 0.50\% in the main paper to the VES 1.35\% at $Re_G \sim 212 \times 10^3$, we observe that the $\langle u^2 \rangle$ for VES 0.50\% is nearly equal to the water results, whereas the VES 1.35\% solution profile at $Re_G \sim 212 \times 10^3$ has lower $\langle u^2 \rangle$ for the entirety of the duct half height. Thus, it appears that increasing $Re_G$ to $212 \times 10^3$ does not result in the same \%DR loss to reach the $\langle u^2 \rangle$ values of water. The work by \citet{li2005} was one of the few that investigated the Reynolds number effect on turbulent drag reduction of a dilute surfactant (CTAT) solution. They found as the Reynolds number increases, \%DR decreases as the Reynolds stresses approach the water values, with agreement to our results. However, even with \%DR $\sim 0$, the turbulent flow statistics were not the same as water at a similar Reynolds number. This happens because, according to their measurements at high Reynolds number, there was still a small contribution of viscoelastic shear stresses to the total shear stress balance that was insufficient to provide significant levels of drag reduction.

The $\langle v^2 \rangle$ and $-\langle uv \rangle$ results in \hyperref[degr-fig2]{Figure~\ref*{degr-fig2}} (b) and (c), respectively, are expected the context seen in the friction factor data of the main paper. Both $\langle v^2 \rangle$ and $-\langle uv \rangle$ values increase as we increase $Re_G$, implying further breakdown of the molecular structure of the VES with larger inertial effects, to the point where we reach rather low levels of \%DR. At the highest velocities investigated in our experiments, we could not reach the values of the VES 0.50\% experiments, and the micellar structure is not as broken down as a result. The trend for the skewness results in \hyperref[degr-fig2]{Figure~\ref*{degr-fig2}} (d) is akin to decreasing the concentration of VES while keeping a constant $Re_G$. As the bulk velocity is increased, the skewness further resembles the water results, in a larger range of positions from centreline to the near the wall.

\begin{figure}
	\centering
	\includegraphics*[width=120mm]{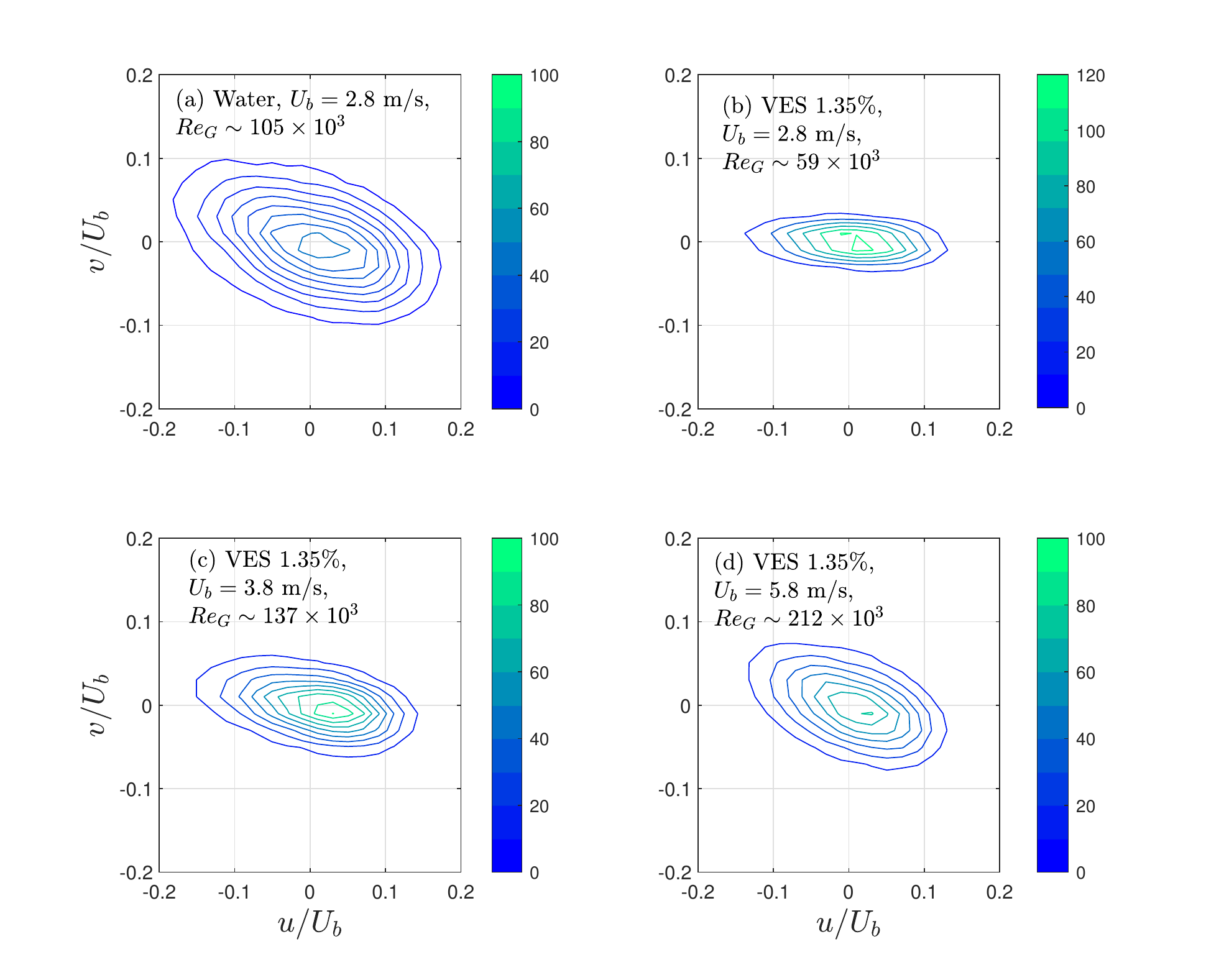}
	\caption{\fontsize{9}{9}\selectfont Joint probability density functions of water (a) and 1.35\% VES solution at $Re_G \sim 59 \times 10^3$ (b), $137 \times 10^3$ (c) and $212\times 10^3$ (d).}
	\label{degr-fig3}
\end{figure}

\begin{figure}
	\centering
	\includegraphics*[width=140mm]{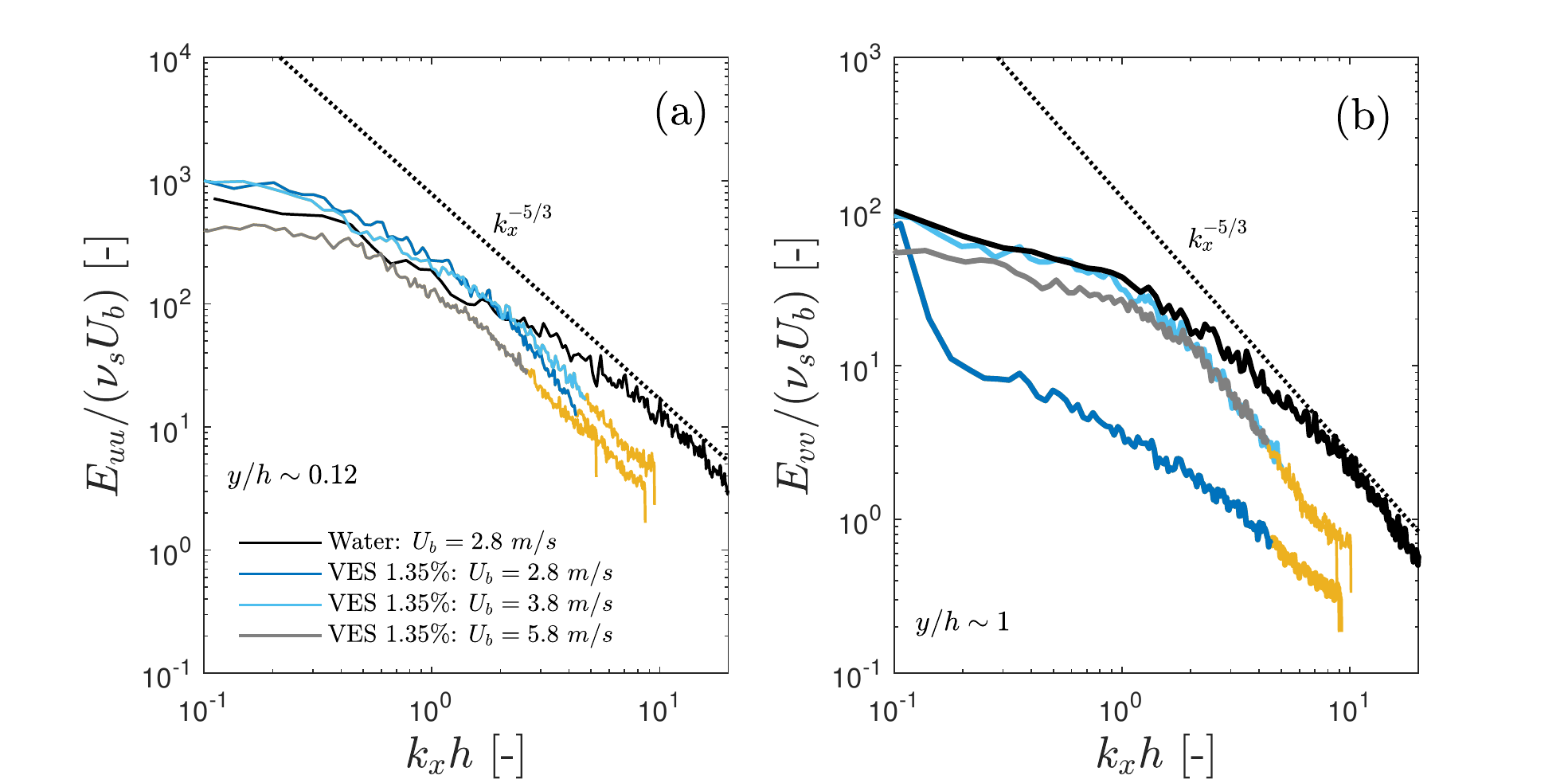}
	\caption{\fontsize{9}{9}\selectfont Streamwise power spectral densities of the 1.35\% VES solution at $U_b = 2.8$, $3.8$ and $5.8~m/s$ at $y/h = 0.12$ (a) and $y/h = 1$ (b). We do not show fits to the $E_{uu}(k_x)$ curves here due to the limited data rate of the experiments. }
	\label{degr-fig4}
\end{figure}

The effect of $Re_G$ on the turbulent structure of the VES 1.35\% solutions is now investigated via joint probability density functions (JPDFs) and power spectral densities (PSDs) in \hyperref[degr-fig3]{Figure~\ref*{degr-fig3}} and \hyperref[degr-fig4]{Figure~\ref*{degr-fig4}}, respectively, at $y/h \sim 0.3$. \hyperref[degr-fig3]{Figure~\ref*{degr-fig3}} (a) shows the JPDF of water and \hyperref[degr-fig3]{Figure~\ref*{degr-fig3}} (b), the JPDF of the VES 1.35\% fluid at MDR, which have been discussed in the main paper. As we increase $Re_G$ in \hyperref[degr-fig3]{Figure~\ref*{degr-fig3}} (c), the shape  of the JPDF of VES 1.35\% further resembles the JPDF of water, with increased ejection (Q2) and sweep (Q4) motions \citep{mohammadtabar2017}, which indicate larger turbulence production, and therefore decreased turbulence attenuation due to a less viscoelastic micellar fluid. Increasing the $Re_G$ in \hyperref[degr-fig3]{Figure~\ref*{degr-fig3}} (d) causes the JPDF to be of  similar shape than the water case in \hyperref[degr-fig3]{Figure~\ref*{degr-fig3}} (a), with higher probabilities of Q2 and Q4 motions near the wall, but with a narrower distribution of $u$ and $v$ fluctuations, because at $Re_G \sim 212 \times 10^3$ the micelles are further broken down and appear similar to the VES 0.50\% solution.

The larger $Re_G$ also affects the energy spectra of streamwise velocity fluctuations $E_{uu}$, as shown in \hyperref[degr-fig4]{Figure~\ref*{degr-fig4}} (a) and (b), at $y/h \sim 0.12$ and $y/h \sim 1$, respectively. We remind the reader that the low data rate of our measurements prevented us from measuring the high-wavenumber range of the spectra. Therefore, we show the cut-off data at a maximum frequency of approximately 1/4 of the total data rate, and also the uncut data in yellow, which could be prone to interpolation errors but are useful at least as a qualitative extrapolation. Near the wall the values of $E_{uu}$ are quite close to each other, which is somewhat surprising given that the $\langle u^2 \rangle$ profiles of the VES at $Re_G \sim 212 \times 10^3$ are quite different from the rest. Unfortunately our data rates do not allow us to further differentiate the power spectra at different $Re_G$. At the centreline ($y/h \sim 1$), the energy quantity in the turbulent core increases by a significant amount, as seen from the $E_{uu}$ results for the VES 1.35\% fluid at $Re_G \sim 59 \times 10^3$ and $137 \times 10^3$. Interestingly, the $E_{uu}$ for $Re_G \sim 137 \times 10^3$ and $212 \times 10^3$ are almost the same in the centreline, which implies that differences in $E_{uu}$ could be more evident in other positions in the duct. Overall, the results show that micellar breakage increases not only the energy quantity in all wavenumbers, but the slope of $E_{uu}$ appears to reach ever closer to the inertial range scale of $k_x^{-5/3}$. This conclusion is in agreement to was observed in the 1D spectra of turbulent kinetic energy by \citet{pereira2018} in DNS of drag-reduced plane Couette flows with degradation.

\section{Convergence of turbulent statistics}

\begin{figure}
\centerline{\includegraphics*[width=140mm]{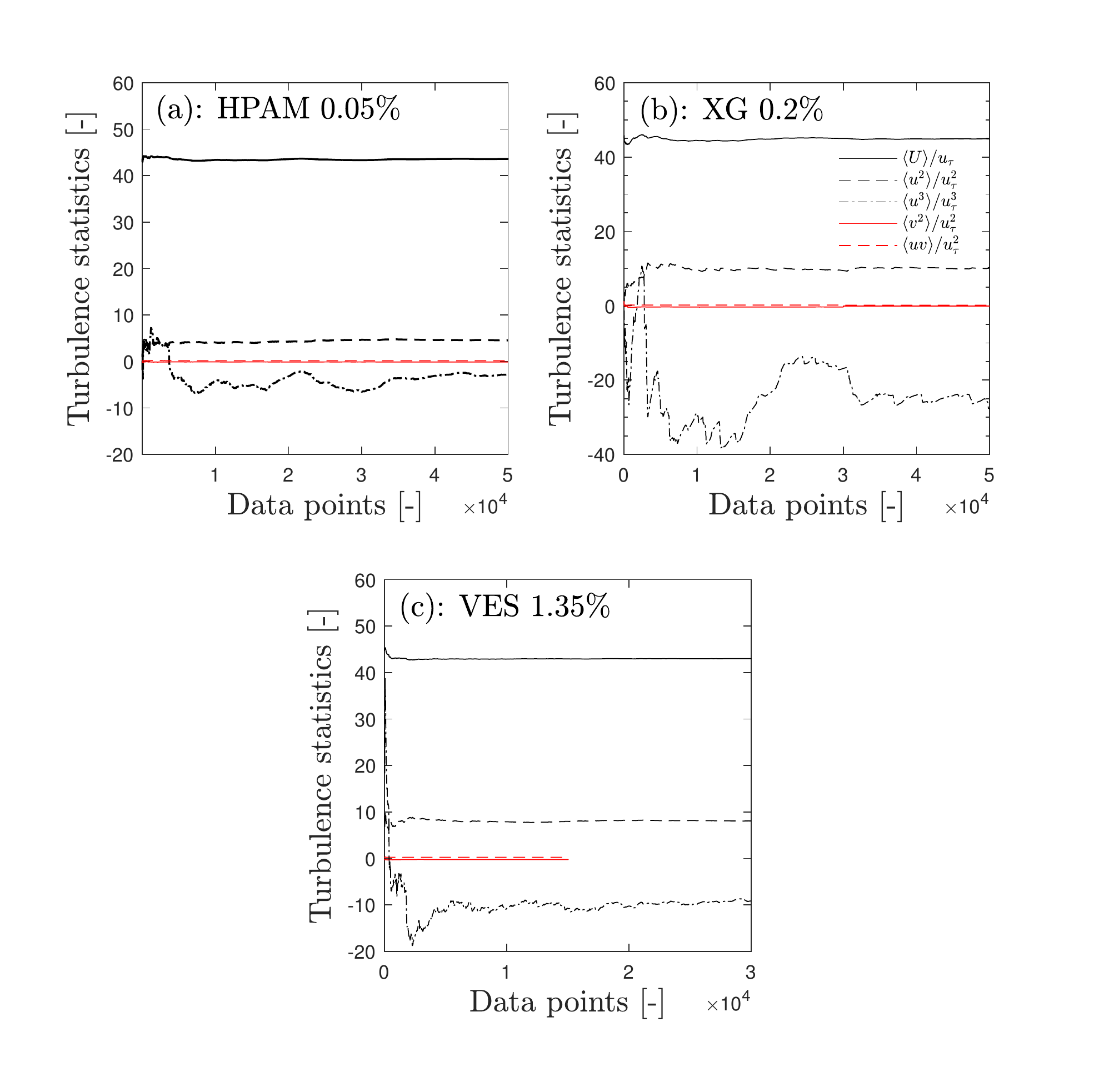}}
\caption{Convergence of turbulence statistics in the turbulent flow of HPAM 0.05\% at $U_b = 3.8~m/s$ (a), XG 0.2\% at $U_b = 3.8~m/s$ (b) and VES 1.35\% at $U_b = 2.8~m/s$, in the position $y/h \sim 0.28$. Data in black were measured in non-coincidence mode, and data in red were measured in coincidence mode.}
\label{sup-fig2}
\end{figure}

We present a verification of convergence of turbulence statistics for HPAM 0.05\%, XG 0.2\% and VES 1.35\% in \hyperref[sup-fig2]{figure~\ref*{sup-fig2}} (a), (b) and (c), respectively. We observe that for all fluids, nearly all turbulence quantities converge with less than 20000 data points, with the excepction of $\langle u^3 \rangle$, which requires over 30000 points to achieve reasonable convergence with the polymer solutions. The coincident mode data for VES was taken at somewhat low data rates, but convergence was very good at 15000 points, as we can see from \hyperref[sup-fig2]{figure~\ref*{sup-fig2}} (c). The VES data from non-coincidence mode showed excellent convergence with 30000 measurement points as well.

\bibliographystyle{elsarticle-num-names}
\bibliography{EDAB_drag_reduction}